\documentclass[a4paper]{jpconf}

\usepackage{graphicx,color,rotating,pifont}
\usepackage{amssymb,bm}
\usepackage{iopams}
\usepackage{ae}
\usepackage{pstricks}
\usepackage{array}
\usepackage[numbers,sort&compress]{natbib}
\usepackage{url}
\expandafter\let\csname equation*\endcsname\relax
\expandafter\let\csname endequation*\endcsname\relax
\usepackage{amsmath}

\allowdisplaybreaks

\DeclareMathSymbol{\NS}{\mathord}{AMSb}{"4E}

\newcommand{\ket}[1]{\ensuremath{\,|{#1}\rangle}}

\newcommand{\matrixe}[3]{\ensuremath{\langle{#1}|\,{#2}\,|{#3}\rangle}}

\newcommand{\comm}[2]{\ensuremath{[{#1},{#2}]}}

\newcommand{\op}[1]{\ensuremath{#1}}
\newcommand{\adj}[1]{\ensuremath{{{#1}}^{\dag}}}

\newcommand{\totd}[2]{\ensuremath{ \frac{d {#1}} {d {#2}} }}

\newcommand{\nord}[1]{\ensuremath{:\!#1:}}

\newcommand{\aO}{\ensuremath{\op{a}}}

\newcommand{\etaO}{\ensuremath{\op{\eta}}}

\newcommand{\aaO}{\ensuremath{\adj{\op{a}}}}

\newcommand{\AO}{\ensuremath{\op{A}}}
\newcommand{\HO}{\ensuremath{\op{H}}}

\newcommand{\TO}{\ensuremath{\op{T}}}
\newcommand{\UO}{\ensuremath{\op{U}}}
\newcommand{\VO}{\ensuremath{\op{V}}}

\newcommand{\UUO}{\ensuremath{\adj{\op{U}}}}

\newcommand{\qOV}{\ensuremath{\vec{\op{q}}}}
\newcommand{\rOV}{\ensuremath{\vec{\op{r}}}}

\newcommand{\sigmaOV}{\ensuremath{\vec{\op{\sigma}}}}

\renewcommand{\AC}{\ensuremath{\mathcal{A}}}

\newcommand{\OC}{\ensuremath{\mathcal{O}}}

\newcommand{\nn}{\ensuremath{\bar{n}}}

\newcommand{\Trel}{\ensuremath{\TO_\text{rel}}}

\newcommand{\Tcm}{\ensuremath{\TO_\text{cm}}}

\newcommand{\Hint}{\ensuremath{\HO_\text{int}}}

\newcommand{\Hd}{\ensuremath{\HO_{d}}}
\newcommand{\Hod}{\ensuremath{\HO_{od}}}

\newcommand{\nuc}[2]{\ensuremath{^{#2}\mathrm{#1}}}

\newcommand{\fm}{\ensuremath{\,\text{fm}}}
\newcommand{\fmi}{\ensuremath{\,\text{fm}^{-1}}}

\newcommand{\keV}{\ensuremath{\,\text{keV}}}
\newcommand{\MeV}{\ensuremath{\,\text{MeV}}}

\newcommand{\lambdaSRG}{\ensuremath{\lambda}}

\newcommand{\NNNLO}{N$^3$LO}
\newcommand{\NNLO}{NNLO}
\newcommand{\NNLOsat}{$\text{NNLO}_\text{sat}$}

\definecolor{FGViolet}{rgb}{0.61,0.32,0.61}
\definecolor{FGDarkBlue}{rgb}{0,0,0.6}
\definecolor{FGBlue}{rgb}{0,0,0.8}
\definecolor{FGLightBlue}{rgb}{0.2, 0.6, 0.8}
\definecolor{FGGreen}{rgb}{0.2,0.7,0.2}
\definecolor{FGLightGreen}{rgb}{0.4,1,0.4}
\definecolor{FGYellow}{rgb}{1,0.95,0}
\definecolor{FGOrange}{rgb}{0.95,0.5,0.1}
\definecolor{FGRed}{rgb}{0.8,0,0}
\definecolor{FGWhite}{rgb}{1,1,1}
\definecolor{FGLightGray}{rgb}{0.8,0.8,0.8}
\definecolor{FGGray}{rgb}{0.5,0.5,0.5}
\definecolor{FGDarkGray}{rgb}{0.3,0.3,0.3}
\definecolor{FGBlack}{rgb}{0,0,0}

\begin{document}
\title{Nuclear Structure from the In-Medium Similarity Renormalization Group}

\author{H.~Hergert$^{1,*}$, J.~M.~Yao$^{2,1}$, T.~D.~Morris$^{3,4}$, N.~M.~Parzuchowski$^{1,5}$, S.~K.~Bogner$^1$, J.~Engel$^2$}

\address{$^1$NSCL/FRIB Laboratory and Department of Physics \& Astronomy, Michigan State University,
East Lansing, MI 48824}

\address{$^2$Department of Physics \& Astronomy, University of North Carolina at Chapel Hill, Chapel Hill, NC 27599}

\address{$^3$Department of Physics \& Astronomy, University of Tennessee-Knoxville, Knoxville, TN 37996}
\address{$^4$Oak Ridge National Laboratory, Oak Ridge, TN 37830}

\address{$^5$Department of Physics, The Ohio State University, Columbus, OH 43210}

\address{$^*$Corresponding author, E-mail: hergert@frib.msu.edu}

\begin{abstract}
Efforts to describe nuclear structure and dynamics from first principles 
have advanced significantly in recent years. Exact methods for light 
nuclei are now able to include continuum degrees of freedom and treat 
structure and reactions on the same footing, and multiple approximate,
computationally efficient many-body methods have been developed that
can be routinely applied for medium-mass nuclei. This has made it possible 
to confront modern nuclear interactions from Chiral Effective Field Theory,
that are rooted in Quantum Chromodynamics with a wealth of experimental data. 

Here, we discuss one of these efficient new many-body methods, the In-Medium 
Similarity Renormalization Group (IMSRG), and its applications in modern
nuclear structure theory. The IMSRG
evolves the nuclear many-body Hamiltonian in second-quantized form through
continuous unitary transformations that can be implemented with polynomial 
computational effort. Through suitably chosen generators, we drive the matrix 
representation of the Hamiltonian in configuration space to specific shapes, 
e.g., to implement a decoupling of low- and high-energy scales, or 
to extract energy eigenvalues for a given nucleus.

We present selected results from Multireference IMSRG (MR-IMSRG) calculations 
of open-shell nuclei, as well as proof-of-principle applications for intrinsically 
deformed medium-mass nuclei. We discuss the successes and prospects of merging 
the (MR-)IMSRG with many-body methods ranging from Configuration Interaction
to the Density Matrix Renormalization Group, with the goal of achieving an
efficient simultaneous description of dynamic and static correlations in
atomic nuclei. 

\end{abstract}




\section{Introduction}

Effective Field Theory (EFT) and Renormalization Group (RG) methods
have become important tools of modern (nuclear) many-body theory.
In the Standard Model, Quantum Chromodynamics (QCD) is the fundamental 
theory of the strong interaction, but the description of nuclear observables 
on the level of quarks and gluons is not feasible, except in the lightest 
few-nucleon systems (see, e.g., \cite{Detmold:2015xw}). Nowadays,
nuclear interactions are derived in chiral EFT, which provides a constructive 
framework and organizational hierarchy not only for nuclear interactions,
but also for the corresponding electroweak currents (see, e.g.,
\cite{Epelbaum:2009ve,Machleidt:2011bh,Epelbaum:2015gf,Entem:2015qf,Machleidt:2016yo,Gezerlis:2014zr,Lynn:2016ec,Lynn:2017eu,Pastore:2009zr,Pastore:2011dq,Piarulli:2013vn,Kolling:2009yq,Kolling:2011bh}). 
Chiral EFT is essentially a low-momentum expansion of nuclear observables, whose 
high-momentum (short-range) physics is not explicitly resolved, but parametrized by the 
so-called low-energy constants (LECs) of the theory. In principle, LECs can be determined 
by matching results of chiral EFT and (Lattice) QCD for observables
accessible by both theories, but such efforts are still in their infancy
\cite{Barnea:2015ns,Contessi:2017ik}. In practice, LECs are therefore fit to
experimental data for low-energy QCD observables, typically in the pion-nucleon ($\pi{}$N), 
two-nucleon (NN), and three-nucleon (3N) sectors
\cite{Epelbaum:2009ve,Machleidt:2011bh,Ekstrom:2015fk,Shirokov:2016wo}.

\begin{figure}[t]
  \begin{center}
    \includegraphics[width=\textwidth]{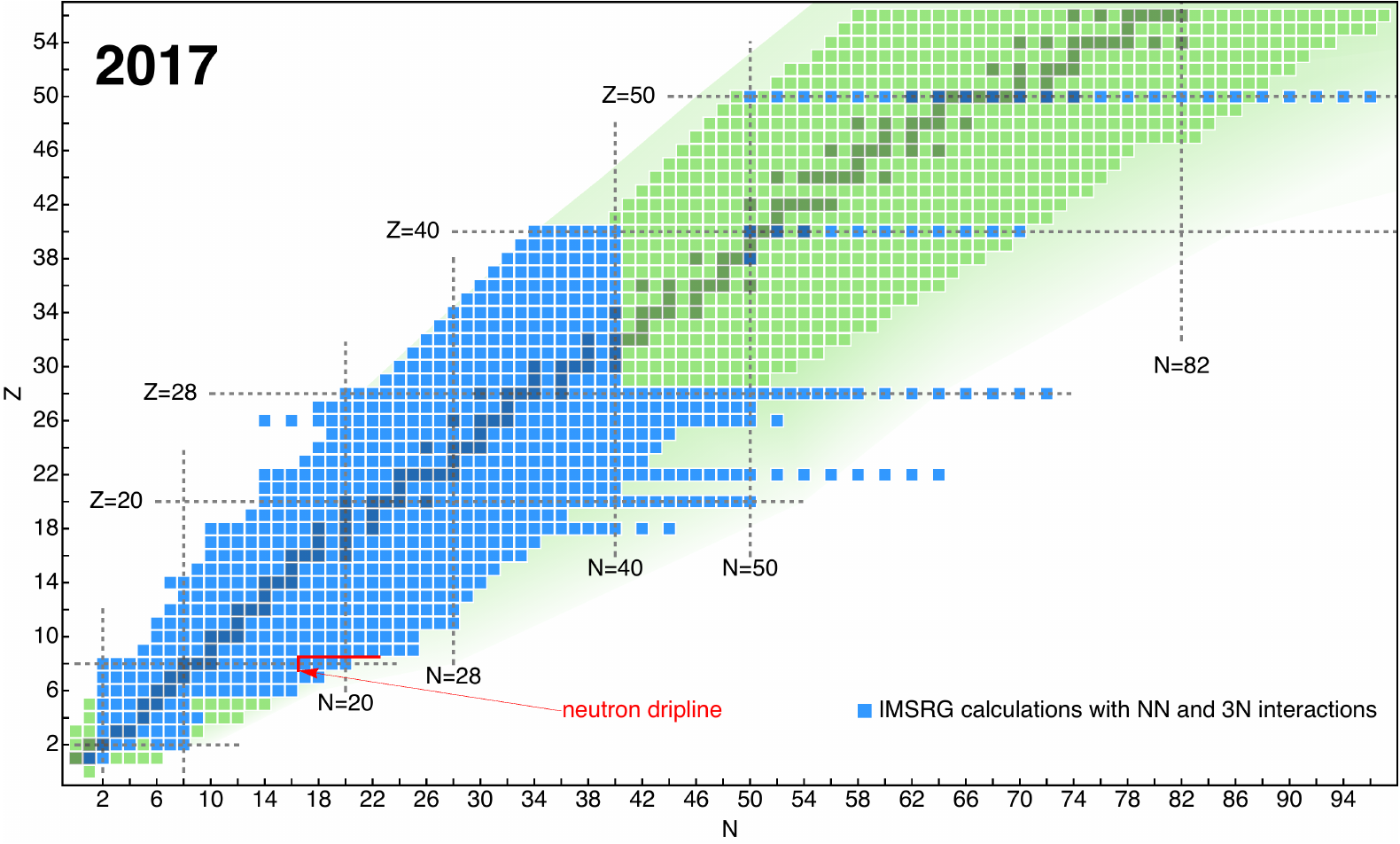}
  \end{center}
  \vspace{-10pt}
  \caption{
    \label{fig:chart2017}
    Survey of IMRSG-based calculations of nuclear ground-state properties and low-lying
    spectra up to 2017.
  }
\end{figure}

RG methods can be used to smoothly connect EFTs with different resolution 
scales and degrees of freedom \cite{Lepage:1989hf,Lepage:1997py}. Since they 
were introduced in low-energy nuclear physics around the start 
of the millennium \cite{Bogner:2003os,Bogner:2007od,Bogner:2010pq,Furnstahl:2013zt}, 
they have greatly enhanced our understanding of the nuclear many-body 
problem by providing a systematic framework for ideas that had been 
discussed in the nuclear structure community since the 1950s. The Similarity 
Renormalization Group (SRG) \cite{Glazek:1993il,Wegner:1994dk}, in particular,
is now widely used to decouple low- and high-momentum physics in nuclear
interactions \cite{Bogner:2006qf,Bogner:2010pq,Roth:2010ys,Tichai:2016vl}. 
This decoupling leads to a greatly improved convergence behavior in many-body
methods that rely on a configuration space, enabling applications of these methods 
to increasingly heavy nuclei
\cite{Barrett:2013oq,Jurgenson:2013fk,Hergert:2013ij,Roth:2014fk,Binder:2014fk,Hagen:2014ve,Hagen:2016rb,Tichai:2016vl,Hergert:2017kx}.
As an example, Fig.~\ref{fig:chart2017} shows the nuclei for which converged 
ground-state calculations have been performed in the In-Medium SRG (IMSRG) 
framework that is the focus of this work, starting from NN and 3N interactions 
derived in chiral EFT. This enhanced reach of \emph{ab initio} nuclear structure 
theory has made it possible to confront chiral interactions
with a wealth of existing nuclear data, with the goal of improving these interactions
in order to reduce discrepancies between theory and experiment.

In the IMSRG, we apply the idea of decoupling energy scales to directly
tackle the nuclear many-body problem \cite{Tsukiyama:2011uq,Hergert:2013mi,Hergert:2016jk,Hergert:2017kx}.
In essence, we use RG flow equations to decouple physics at different excitation 
energy scales of the nucleus, and render the Hamiltonian matrix in 
configuration space block or band diagonal in the process. With an appropriately 
chosen decoupling strategy, it is even possible to extract specific eigenvalues 
and eigenstates of the Hamiltonian, and thereby solve the quantum many-body problem.
We will see that this can be achieved on the operator level, without actually 
constructing the Hamiltonian matrix in a factorially growing configuration
space basis. We will also discuss how the IMSRG evolution can be viewed as 
a re-organization of the many-body expansion in which correlations are 
absorbed into an \emph{RG-improved} Hamiltonian. 
The idea of using Hamiltonian flow equations to solve quantum many-body problems
was already discussed in Wegner's initial SRG work \cite{Wegner:1994dk} 
(also see \cite{Kehrein:2006kx} and references therein), and it is used 
in a variety of flavors in domains like solid-state physics 
\cite{Heidbrink:2002kx,Drescher:2011kx,Krull:2012bs,Fauseweh:2013zv,Krones:2015ft},
or quantum chemistry \cite{White:2002fk,Yanai:2007kx,Nakatsuji:1976yq,Mukherjee:2001uq,Mazziotti:2006fk,Evangelista:2014rq}.

In the following, we will briefly discuss the salient elements of 
the chiral EFT description of the strong interaction at low energies (Sec.~\ref{sec:eft}),
and the use of SRG evolutions to dial the resolution scale of nuclear
interactions (Sec.~\ref{sec:srg}). We will then describe the IMSRG
formalism and its extension to correlated reference states, with the
specific goal of tackling open-shell systems with multi-reference 
character (Sec.~\ref{sec:mrimsrg}). Selected applications of this
Multireference IMSRG (MR-IMSRG) are presented in Sec.~\ref{sec:app},
followed by a discussion of how the MR-IMSRG can help us tackle 
collective correlations in nuclei (Sec.~\ref{sec:static}). Finally,
we will touch upon the use of IMSRG-improved Hamiltonians in
other many-body methods, and possible interfaces with White's 
Density Matrix Renormalization Group (DMRG) \cite{White:1992qf}, as well as tensor RG
methods for many-body systems on spatial lattices \cite{Vidal:2007if,Evenbly:2009cy,Evenbly:2015ht} 
(Sec.~\ref{sec:heff}). 

\section{\label{sec:eft}Effective Field Theories of the Strong Interaction}

Quantum Chromodynamics (QCD) is the fundamental theory of the strong interaction
between quarks and gluons. One of its essential features is that the strength of
the interaction increases in the low-energy domain that is most relevant for
the structure and reactions of atomic nuclei \cite{Gross:1973pd,Politzer:1973lq}. 
This makes the description of all but the lightest nuclei at the QCD level inefficient 
at best, and impossible at worst. However, strongly interacting matter undergoes a 
phase transition from a 
quark-gluon plasma to a hadronic phase in which quarks are confined in composite 
particles like nucleons and pions. These particles can be used as building 
blocks for an EFT of the strong interaction.

Following Weinberg \cite{Weinberg:1991rw,Weinberg:1996bb}, the dynamics of an
EFT are governed by a Lagrangian consisting of all interactions that are consistent
with the symmetries of the underlying theory. Of particular importance is the chiral 
symmetry of QCD: massless QCD is invariant under independent rotations of left- and 
right-handed quarks (hence the name chiral symmetry), expressed by the symmetry group
$SU(2)_L\times SU(2)_R$ \cite{Weinberg:1991rw,Epelbaum:2010nr,Machleidt:2011bh}. This 
symmetry is explicitly broken by the quark mass terms,
but since the masses of the up and down quarks are small, it is still very relevant for
the interactions of nucleons and pions, which are built from those light quarks. Since
the symmetry group $SU(2)_L\times SU(2)_R$ can be equivalently represented as a product
of vector and axial vector rotations, $SU(2)_V\times SU(2)_A$, chiral symmetry would
imply the existence of parity doublets of strongly interacting particles which are
not found in nature. Thus, we can conclude that chiral symmetry must be 
broken spontaneously, and it turns out that it is the axial component of the theory
that undergoes this symmetry breaking. The invariance of nuclear and nucleon-pion
interactions under the remaining $SU(2)_V$ group is the well-known nuclear isospin symmetry \cite{Heisenberg:1932ph}.
The relevant particle multiplets for chiral EFT are the proton and neutron, as well
as the $\pi_+,\pi_0, \pi_-$ mesons. The spontaneous breaking of chiral symmetry implies the existence of 
massless Goldstone bosons \cite{Goldstone:1961oq}, which can be readily identified 
with the pions.

\begin{figure}[t]
  \begin{center}
    \includegraphics[width=0.9\textwidth]{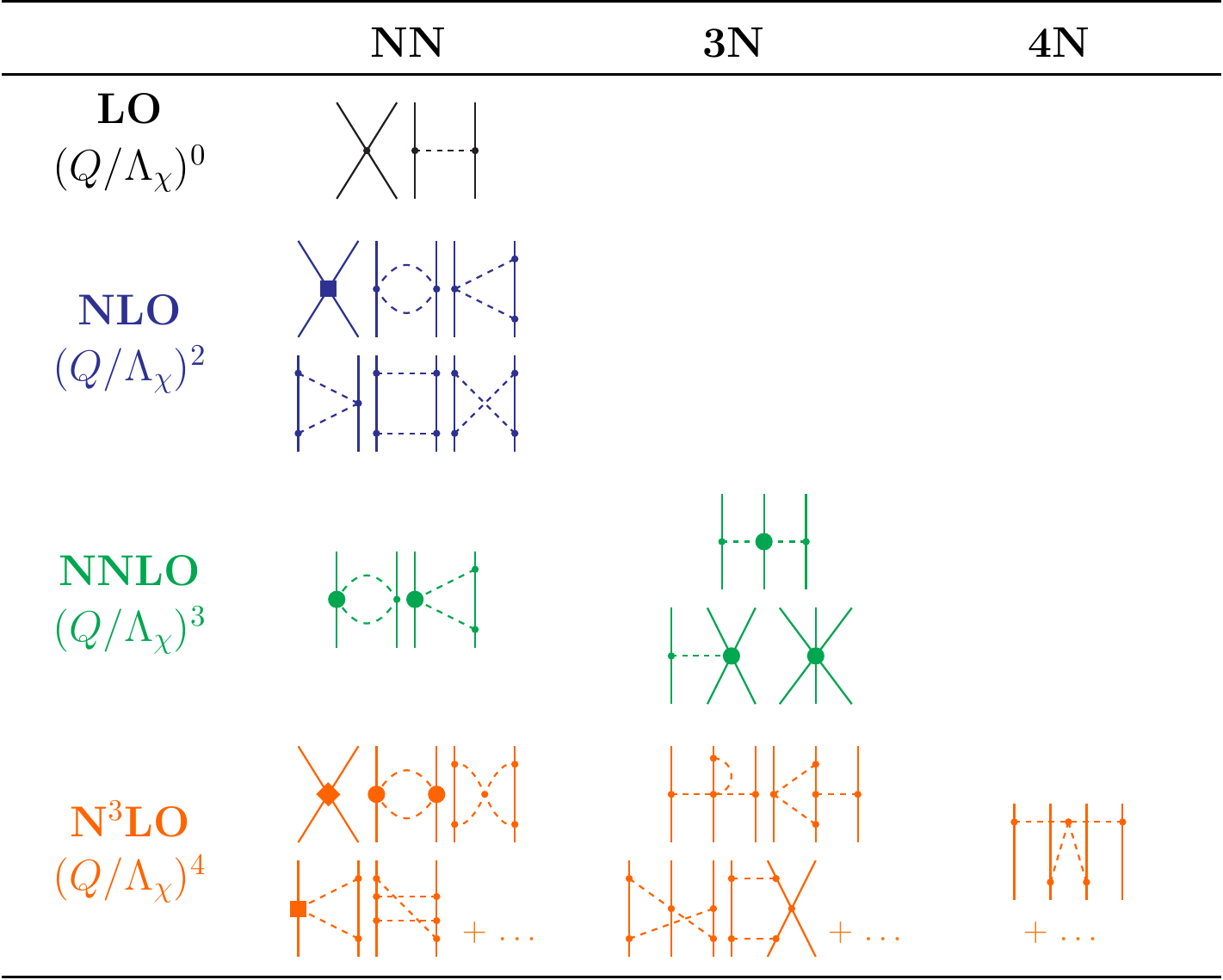}
  \end{center}
  \caption{\label{fig:chiral}Chiral two-, three- and four nucleon forces 
  through next-to-next-to-next-to-leading order (\NNNLO{}) (see, e.g., 
  \cite{Epelbaum:2009ve,Machleidt:2011bh,Machleidt:2016yo}). Dashed lines
  represent pion exchanges between nucleons. The large solid circles, boxes
  and diamonds represent vertices that are proportional to low-energy constants (LECs) of
  the theory (see text).}
\end{figure}

Another important consequence of chiral symmetry and its breaking is that 
only derivatives of the pion fields couple to the nucleon. Thus,
interaction vertices (and Feynman diagrams) become proportional to ($q/\Lambda_\chi$),
where $q\sim m_\pi \sim k_F\approx 140\,\MeV/c$ is the typical momentum of a system
of nucleons and pions, and $\Lambda_\chi$ is the so-called breakdown scale 
of the EFT. The breakdown scale is associated with 
physics that is not explicitly taken into account by the constructed EFT. In 
chiral EFT, $\Lambda_\chi$ is traditionally considered to be in the range 
$700-1000\,\MeV$, with the lower value corresponding to the mass of the $\rho$ meson 
that is not an explicit degree of freedom of the theory. Assumptions about 
$\Lambda_\chi$ have come under increasing scrutiny as
applications in medium-mass nuclei have revealed shortcomings of the chiral
interactions (see, e.g., \cite{Melendez:2017fj} and references therein). 

Nevertheless, the proportionality of the interaction vertices to $q/\Lambda_\chi<1$
makes it possible to organize the chiral Lagrangian in powers of this ratio.
Based on this power-counting scheme, one can then develop a systematic 
low-momentum expansion of nuclear interactions, as shown in Fig.~\ref{fig:chiral} 
(see \cite{Weinberg:1991rw,Epelbaum:2009ve,Machleidt:2011bh,Machleidt:2016yo}).
These interactions consist of pion exchanges between nucleons, indicated
by dashed lines, as well as nucleon contact interactions. Importantly,
the diagrams shown in Fig.~\ref{fig:chiral} contain different types of
pion-nucleon, nucleon-nucleon, and three-nucleon vertices that are proportional
to specific LECs of chiral EFT. As mentioned above, these LECs encode 
physics that is not explicitly resolved, either because it corresponds to a high
momentum scale, or involves degrees of freedom that are not explicitly treated
by the theory. Eventually, one hopes to calculate these LECs directly from
the underlying QCD, i.e., through matching results from the nuclear sector
to Lattice QCD calculations (see \cite{Barnea:2015ns,Contessi:2017ik} for 
first attempts in that direction). In the mean time, the LECs are fit to experimental
data.

The organization of nuclear forces from leading (LO) to next-to-next-to-next-to-leading 
order (\NNNLO{}) and beyond in the chiral expansion is appealing for a number of reasons.
By starting from a chiral Lagrangian, we obtain a \emph{consistent} set of
two-, three- and higher many-nucleon interactions, and the power counting
explains the empirical hierarchy of these nuclear forces, i.e., $V_\text{NN}>V_\text{3N}>V_\text{4N}>\ldots$.
Moreover, the chiral Lagrangian can be readily extended with couplings to the
electroweak sector by replacing the derivatives with their gauge covariant
versions. This implies that the same LECs appear in the nuclear interactions 
and electroweak currents, and the LECs can therefore be constrained by electroweak 
observables (see, e.g., \cite{Gazit:2009qf,Pastore:2011dq,Kolling:2011bh,Piarulli:2013vn}).
Last but not least, by truncating the chiral expansion at a given order,
we will eventually be able to quantify the theoretical uncertainties of 
the interactions and operators that serve as input for nuclear many-body 
theory, once some persistent issues are resolved (see, e.g.,  
\cite{Machleidt:2016yo,Reinert:2017fv,Lynn:2016ec,Lynn:2017eu,Valderrama:2017hb,Sanchez:2018pi}).

\section{\label{sec:srg}The Similarity Renormalization Group}
As mentioned in the introduction, RG methods are natural companions for EFTs, 
because they make it possible to smoothly dial the resolution scale of those theories.
In low-energy nuclear physics, RG methods have played an important role in
formalizing ideas on the renormalization of nuclear interactions and many-body
effects that had been discussed by the community for a long time (see \cite{Bogner:2003os,
Bogner:2007od,Bogner:2010pq,Furnstahl:2013zt} and references therein). 
For instance, local NN interactions with repulsive cores at short inter-particle 
distances or non-local interactions that reproduce NN scattering data equally well, 
can yield significantly different properties for nuclei and nuclear matter (see, e.g.,
\cite{Coester:1970tt,Bethe:1971qf}). Colloquially, interactions with strong
short-range repulsion are referred to as ``hard'' interactions because they
induce strong many-body correlations that make the nuclear many-body problem
non-perturbative. Coester and co-workers \cite{Coester:1970tt} showed that nuclear 
interactions can be softened via unitary transformations that shift strength
from the repulsive core into non-local contributions, thereby making mean-field
and finite-order beyond mean-field approximations to the many-body problem
viable. While their work and similar investigations explicitly recognized 
that such transformations 
lead to induced many-body forces, those forces were neglected, and the discussion
very much revolved around using these techniques to pin down the ``true'' 
nucleon-nucleon potential.

From the modern RG perspective, there is no such thing as a true potential,
but rather infinitely many unitarily equivalent representations of low-energy
QCD. In this context, the use of soft or hard NN interactions is related to
the choice of a low or high resolution scale for the description of the nuclear
many-body system. When an RG is used to dial the resolution, the induced many-body 
forces \emph{must} be accounted for  
so that observables remain invariant under the RG flow (see below). For instance,
induced 3N forces play a crucial role in obtaining proper nuclear matter saturation 
properties if we work at low resolution scale (see, e.g., \cite{Kuckei:2003lk,Hebeler:2011dq,Hebeler:2013qa}).
Of course, 3N (and higher) forces are present anyway in an EFT treatment of the strong 
interaction, as discussed in the previous section.

\subsection{SRG Evolution of Nuclear Forces}

\begin{figure*}[t]
  \begin{center}
    \includegraphics[width=0.9\textwidth]{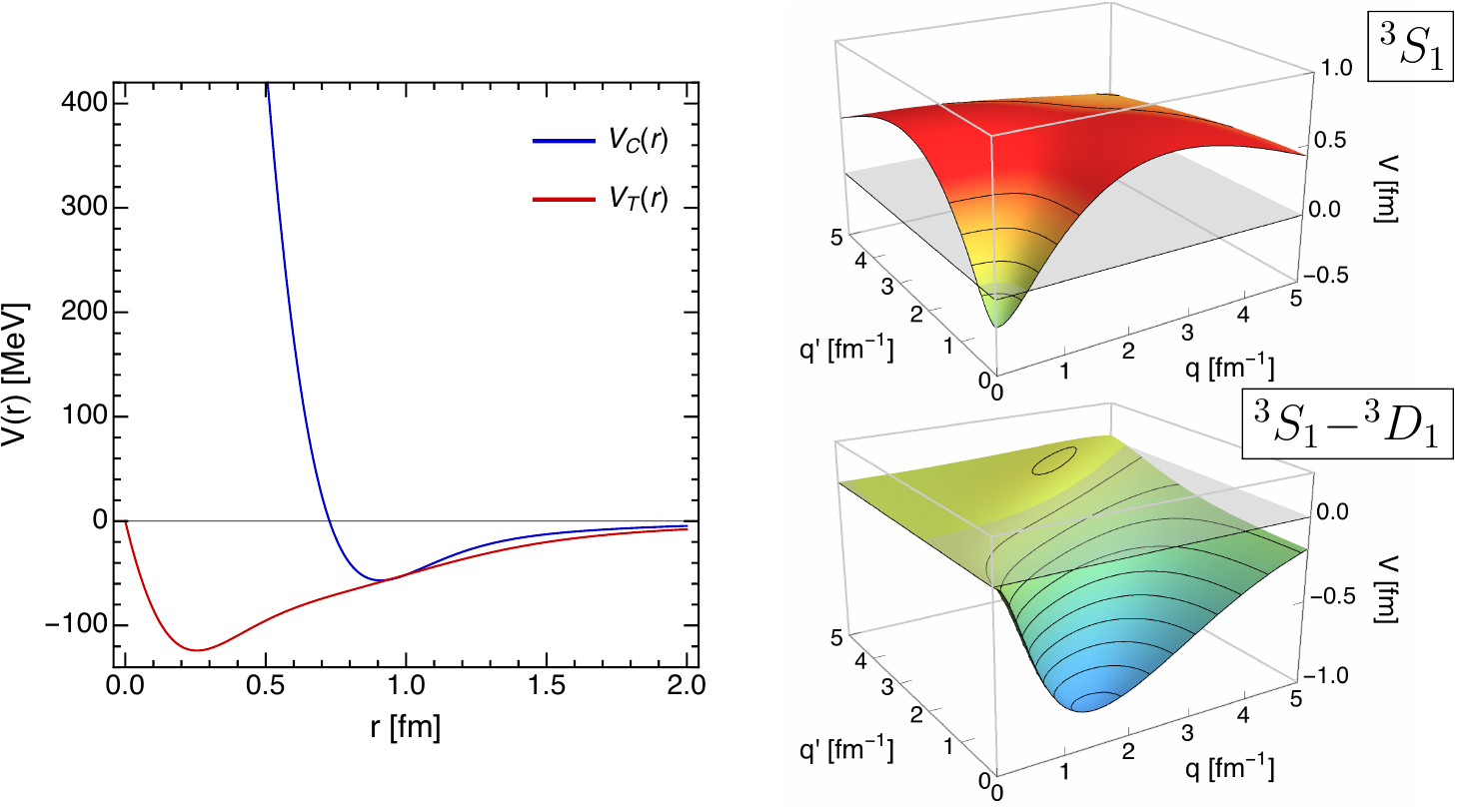}
  \end{center}
  \caption{\label{fig:av18}
    Repulsive core  and tensor force of the Argonne
    V18 NN interaction \cite{Wiringa:1995or} in the $(S,T)=(1,0)$
    channel. In the left panel, the radial dependencies of the central ($V_C(r)$)  and tensor 
    components ($V_T(r)$) of Argonne V18 are shown, while the right panel shows its momentum 
    space matrix elements in the deuteron partial waves.
  }
\end{figure*}

In Fig.~\ref{fig:av18} we show features of the central and tensor forces of the 
Argonne V18 (AV18) interaction \cite{Wiringa:1995or} in the two-body spin-isospin 
channel $(S,T)=(1,0)$, which contains the deuteron bound state. AV18 is a so-called realistic 
interaction that describes NN scattering data with high accuracy, but precedes 
the modern chiral forces discussed above. It is specifically tailored to be as
local as possible, to make it a suitable input for Quantum Monte Carlo calculations 
\cite{Carlson:2015lq,Gezerlis:2014zr,Lynn:2016ec,Lynn:2017eu}. This is the reason
why AV18 has a strong repulsive core in the central part of the interaction. Like 
all NN interactions, it also has a tensor interaction resulting from pion exchange that
mixes states with different orbital angular momenta,
\begin{equation}
  V_T = v_T(r)S_{12}(r)\equiv \frac{v_T(r)}{r^2}\left(3(\sigmaOV_1\cdot\rOV)(\sigmaOV_2\cdot\rOV)-(\sigmaOV_1\cdot\sigmaOV_2)r^2\right),
\end{equation}
The radial dependencies of AV18's central and tensor interactions are shown in the left panel 
of Fig.~\ref{fig:av18}. In the corresponding momentum representation, the ${}^3S_1$ partial wave\footnote{We use the conventional partial wave
notation ${}^{2S+1}L_J$, where $L=0,1,2,\ldots$ is indicated by the letters 
$S,P,D,\ldots$. The isospin channel is fixed by requiring the antisymmetry
of the NN wave function, leading to the condition $(-1)^{L+S+T}=-1$.} has 
strong off-diagonal matrix elements, with tails extending to relative momenta 
as high as $|\qOV| \sim 20\,\fmi$. The matrix elements of the ${}^3S_1-{}^3D_1$ 
mixed partial wave, which are generated exclusively by the tensor force, are 
sizable as well. The strong coupling between states with low and high relative 
momenta forces us to use large Hilbert spaces in few- and many-body calculations, 
even if we are only interested in the lowest eigenstates. Consequently, the 
eigenvalues and eigenstates of the nuclear Hamiltonian converge very slowly
with respect to the basis size of the Hilbert space (see, e.g., 
\cite{Barrett:2013oq}). To solve this problem, we can perform an RG evolution of
the NN interaction.

First systematic applications of RG methods for the renormalization of nuclear
interactions were based on RG decimation, i.e., the \emph{projection} of the 
Hamiltonian onto a low-momentum (low-energy) subspace (see \cite{Bogner:2003os,Bogner:2010pq}
and references therein). In recent years, however, the so-called Similarity Renormalization 
Group (SRG) has become the RG method of choice in nuclear physics \cite{Glazek:1993il,Wegner:1994dk,Bogner:2010pq}.
The SRG is based on \emph{unitary transformations} rather than projection methods,
which greatly facilitates the construction of observables that are consistent
with the renormalized interaction.

The basic concept of the SRG is more general than required by RG theory: We aim to simplify 
the structure of the Hamiltonian in a suitable representation through a continuous 
unitary transformation, 
\begin{equation}\label{eq:cut}
  \HO(s)=\UO(s)\HO(0)\UUO(s)\,.
\end{equation}
Here, $H(s=0)$ is the starting Hamiltonian, and the flow parameter $s$ parameterizes 
the unitary transformation. To implement this transformation, we take the derivative 
of Eq.~\eqref{eq:cut} to obtain the operator flow equation
\begin{equation}\label{eq:opflow}
  \totd{}{s}\HO(s) = \comm{\etaO(s)}{\HO(s)}\,,
\end{equation} 
where the anti-Hermitian generator $\etaO(s)$ is related to $\UO(s)$ by
\begin{equation}
  \eta(s)=\totd{U(s)}{s}U^{\dag}(s) = -\eta^{\dag}(s)\,.
\end{equation}
We can choose $\etaO(s)$ to transform the Hamiltonian to almost arbitrary structures 
as we integrate the flow equation \eqref{eq:opflow} for $s\to\infty$. Wegner \cite{Wegner:1994dk} 
proposed the generator
\begin{equation}\label{eq:def_Wegner_general}
  \etaO(s) \equiv \comm{H_d(s)}{H_{od}(s)}\,,
\end{equation}
which is constructed by splitting the Hamiltonian into suitably chosen \emph{diagonal} 
($H_d(s)$) and \emph{off-diagonal} ($H_{od}(s)$) parts. The generator \eqref{eq:def_Wegner_general} 
will then suppress $H_{od}(s)$ as the Hamiltonian is evolved via equation \eqref{eq:opflow} 
(see \cite{Wegner:1994dk,Kehrein:2006kx,Bogner:2010pq,Hergert:2016jk,Hergert:2017kx}). 
Here, the label diagonal merely refers to a desired structure that the Hamiltonian will 
assume in the limit $s\to\infty$, and not strict diagonality. We can make contact with 
renormalization ideas by imposing a block or band-diagonal structure on the Hamiltonian
that implies a decoupling of momentum or energy scales in the RG sense.

\begin{figure*}[t]
  \begin{center}
    \includegraphics[width=0.9\textwidth]{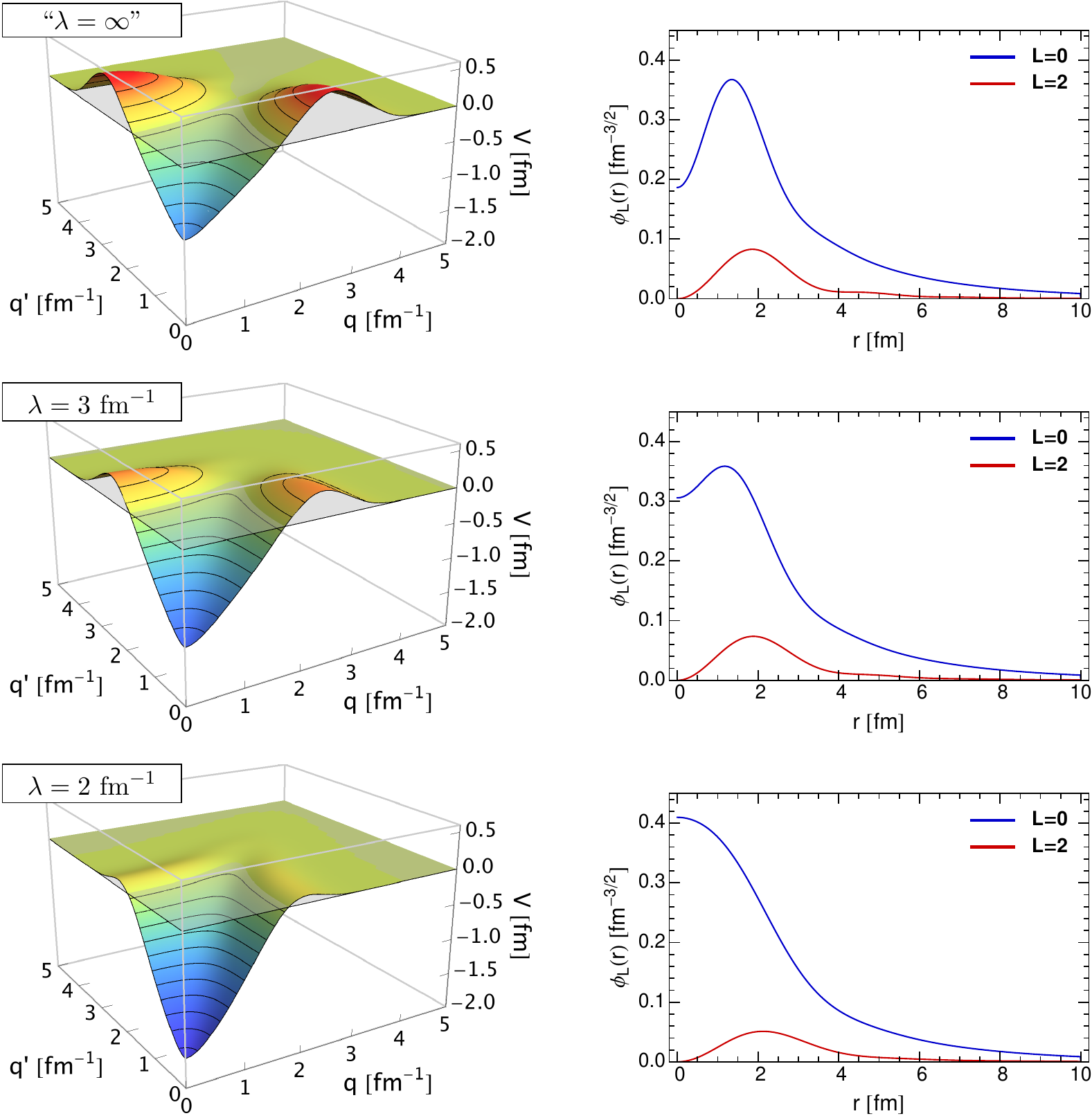}
  \end{center}
  \caption{\label{fig:vsrg_momentum}SRG evolution of the chiral \NNNLO{} NN interaction
  by Entem and Machleidt \cite{Entem:2003th,Machleidt:2011bh}. 
  In the left column, we show the
  momentum-space matrix elements of the interaction in the ${}^3S_1$ partial wave for different values
  of the SRG resolution scale $\lambdaSRG$. The top-most row shows the initial interaction at $s=0\,\fm^4$\,,
  i.e., ``$\lambda=\infty$''. In the right column, we show the $S-$ and $D-$wave components
  of the deuteron wave function that is obtained by solving the Schr\"odinger equation with the corresponding
  SRG-evolved interaction.}
\end{figure*}

As an example of a realistic application, Fig.~\ref{fig:vsrg_momentum} shows the 
SRG evolution of a chiral \NNNLO{} NN interaction. We use a Wegner-type generator 
built from the relative kinetic energy in the two-nucleon system to drive the 
Hamiltonian towards a band-diagonal structure in momentum space:
\begin{equation}\label{eq:def_srg_generator}
  \eta(\lambda) \equiv \Big[\,\frac{\qOV^2}{2\mu}, v(\lambda)\Big]\,.
\end{equation}
We have changed variables from the flow parameter $s$ to $\lambdaSRG=s^{-1/4}$, which has 
the dimensions of a momentum (in natural units), for reasons that will become clear
shortly.

We start from the matrix elements of the initial interaction in the ${}^3S_1$ partial 
wave that are shown in the top row. The interaction has no appreciable strength in states 
with momentum differences $|\Delta\qOV|=|\qOV'-\qOV|\gtrsim 4.5\,\fmi$. 
By evolving the initial interaction to $\lambdaSRG=3\,\fmi$ and then to $2\,\fmi$, the 
off-diagonal matrix elements are suppressed, and the interaction is almost entirely 
contained in a block of states with $|\Delta\qOV|\sim 2\,\fmi$, except for a weak 
diagonal ridge. Thus, we see that $\lambdaSRG$ limits the momentum that the interaction
can transfer between pairs of nucleons, and we identify $\lambdaSRG$ with the resolution 
scale of the evolved interaction.

Next to the matrix elements, we show deuteron wave functions obtained by solving the 
Schr\"odinger equation with Hamiltonian at each $\lambdaSRG$. For the initial 
interaction, the $S-$wave ($L=0$) component of the wave function is suppressed at 
small relative distances, which reflects short-range correlations between the nucleons. 
There is also a significant $D-$wave ($L=2$) admixture due to the tensor interaction. 
As we lower the resolution scale, the ``correlation hole'' in the wave function is 
eliminated, and the $D-$wave admixture is reduced because the evolution suppresses 
the ${}^3S_1-{}^3D_1$ matrix elements that cause this mixing \cite{Bogner:2010pq}. 
The structure of the wave function becomes very simple, reminiscent of a Gaussian
ansatz for a pair of independent nucleons. 

\subsection{Induced Forces}

The RG decoupling of low- and high-lying momenta prevents the Hamiltonian from scattering 
nucleon pairs from low- to high-momentum states, which makes it possible to converge 
many-body calculations in much smaller Hilbert spaces than for the harder initial 
interactions \cite{Roth:2011kx,Barrett:2013oq,Jurgenson:2013fk,Roth:2014fk,Hergert:2013ij,Hergert:2013mi,Hergert:2014vn,Hergert:2016jk,Hagen:2010uq,Roth:2012qf,Binder:2013zr,Binder:2014fk,Soma:2011vn,Soma:2013ys,Soma:2014fu,Soma:2014eu}. However, this improvement comes 
at a cost, which is best illustrated by considering the Hamiltonian in second-quantized 
form, assuming only a two-nucleon interaction for simplicity:
\begin{equation}\label{eq:H}
  \Hint = \Trel + V = \frac{1}{4}\sum_{pqrs} \matrixe{pq}{\frac{\qOV_{12}^{\,2}}{2\mu}+v_{12}}{rs}\aaO_p\aaO_q\aO_s\aO_r\,.
\end{equation}
Inserting $\Trel$ and $V$ into Eqs.~\eqref{eq:def_srg_generator} and 
\eqref{eq:opflow}, we obtain
\begin{equation}\label{eq:comm2B_vac}
  \comm{\aaO_i\aaO_j\aO_l\aO_k}{\aaO_p\aaO_q\aO_s\aO_r}=
  \delta_{lp}\aaO_i\aaO_j\aaO_q\aO_s\aO_r\aO_k + \{\aaO\aaO\aaO\aO\aO\aO\} 
  -\delta_{lp}\delta_{kq}\aaO_i\aaO_j\aO_s\aO_r + \{\aaO\aaO\aO\aO\}\,, 
\end{equation}
where the terms in brackets schematically stand for additional two- and three-body 
operators. Thus, even if we start from a pure two-body interaction, the SRG 
flow will induce operators of higher rank. 

\begin{figure*}[t]
  \begin{center}
    \includegraphics[width=0.5\textwidth]{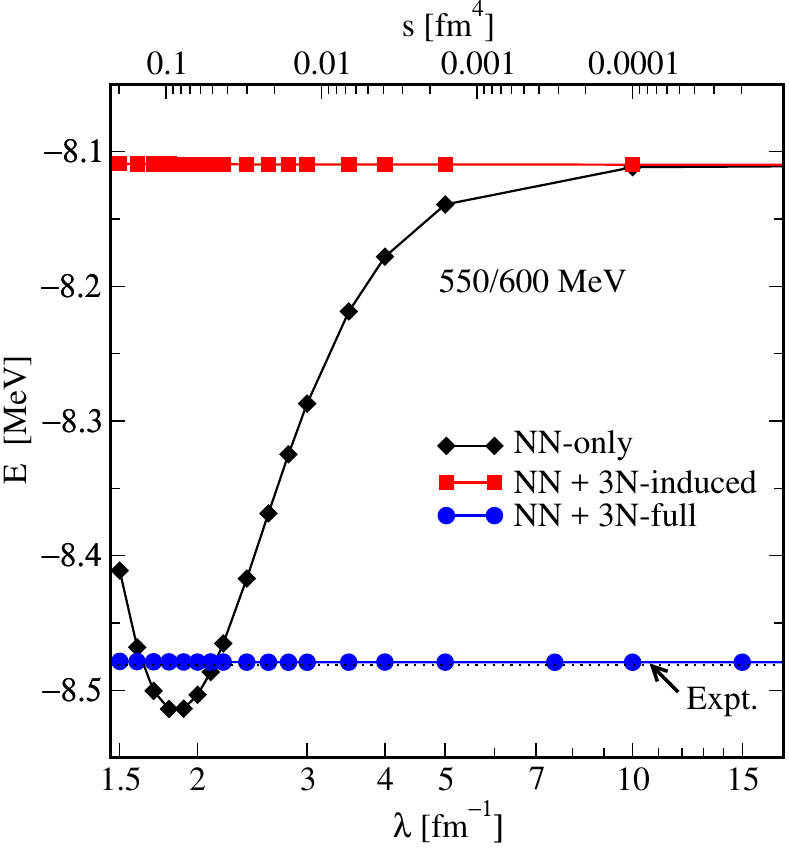}
  \end{center}
  \caption{\label{fig:triton}Ground state energy of $\nuc{H}{3}$ as a function 
  of the flow parameter $\lambdaSRG$ for chiral \NNLO{} $NN$ and $NN\!+\!3N$ 
  interactions (see \cite{Hebeler:2012ly} for details). $NN$-only means initial and 
  induced $3N$ interactions are discarded, $NN\!+\!3N$-induced takes only 
  induced $3N$ interactions into account, and $3N$-full contains initial
  $3N$ interactions as well. The black dotted line shows the experimental 
  binding energy \cite{Wang:2012uq}. Data for the figure courtesy of 
  K.~Hebeler.}
\end{figure*}

Nowadays, state-of-the-art SRG evolutions of nuclear interactions are 
performed in the three-body system \cite{Jurgenson:2009bs,Jurgenson:2011zr,Jurgenson:2013fk, 
Hebeler:2012ly,Wendt:2013uq}. In Fig.~\ref{fig:triton}, we show 
$\nuc{H}{3}$ ground-state energies that have been calculated with
a family of SRG-evolved interactions that is generated from chiral
NN and NN+3N interactions (see \cite{Hebeler:2012ly} for full details). 
If we truncate the evolved interaction and the SRG generator at the two-body 
level (\emph{NN-only}), the SRG evolution is not unitary in the 3N system 
and the energy becomes $\lambdaSRG$ dependent. If we truncate the operators 
at the three-body level instead, induced 3N interactions are properly 
included and unitarity is restored (\emph{NN+3N-induced}): 
The energy does not change as $\lambdaSRG$ is varied. Finally, for the 
curve \emph{NN+3N-full} a 3N force is included in the initial Hamiltonian 
and evolved consistently. The resulting triton ground-state energy is invariant 
under the SRG flow, and it closely reproduces the experimental value 
because the 3N interaction's LECs are fit accordingly 
\cite{Epelbaum:2009ve,Machleidt:2011bh,Gazit:2009qf}.

The triton example shows that it is important to track induced interactions,
especially if we want to use evolved nuclear Hamiltonians as input for 
medium-mass or heavy nuclei rather than the few-nucleon system considered
so far. The nature of the SRG works at least somewhat in our favor, because 
truncations of the SRG flow equations lead to a violation of unitarity that 
manifests as a (residual) dependence of observables on the resolution 
scale $\lambdaSRG$. We can use this dependence as a tool to assess 
the size of missing contributions, although one has to take great
care to disentangle them from the effects of many-body truncations 
\cite{Bogner:2010pq,Jurgenson:2009bs,Hebeler:2012ly,Roth:2011kx,Hergert:2013mi,Hergert:2013ij,Binder:2014fk,
Soma:2014eu,Griesshammer:2015dp}. The empirical observation that SRG 
evolutions to resolutions as low as $\lambdaSRG\sim1.5\fmi$
appear to preserve the natural hierarchy of nuclear forces, i.e.,  
$V_\text{NN} > V_\text{3N} > V_\text{4N} > \ldots$, suggests that we 
can truncate induced forces whose contributions would be smaller than 
the desired accuracy of our calculations. 

\section{\label{sec:mrimsrg}The In-Medium SRG}
\subsection{Concept}
As discussed in Sec.~\ref{sec:srg}, the concept of the SRG is actually
quite general: We can use continuous unitary transformations to drive
a given Hamiltonian (or generic operator) to a desired structure, defined by our choice of the diagonal and off-diagonal
parts of the Hamiltonian, $\Hd$ and $\Hod$. If we choose $\Hd$ to be the
diagonal entries of a system's Hamiltonian matrix, the SRG would amount 
to solving the eigenvalue problem! If we consider applying
this idea in genuine many-body systems like medium-mass or heavy nuclei,
we quickly realize that an SRG evolution of the many-body Hamiltonian
matrix would be a very inefficient approach for solving the Schr\"odinger
equation, because the RHS of Eq.~\eqref{eq:opflow} forces us to repeatedly
perform products of matrices whose dimension grows factorially with the
number of particles and single-particle basis states. Moreover, we would
essentially be working to obtain the complete spectrum of that 
large matrix representation rather than just a comparably small number of
low-lying states that we are truly interested in. 

\begin{figure}[t]
  \begin{center}
    \includegraphics[width=0.9\textwidth]{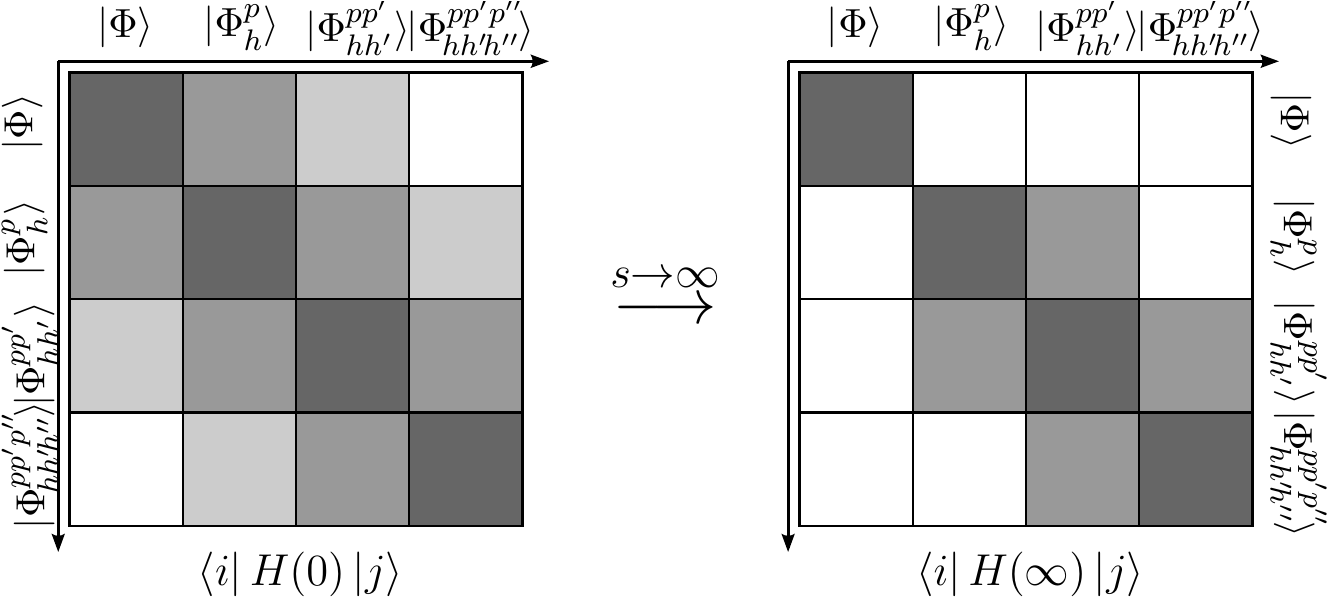}
  \end{center}
  \caption{\label{fig:sr_decoupling}
  Schematic view of IMSRG decoupling in a many-body Hilbert space spanned by
  a Slater determinant reference $\ket{\Phi}$ and its particle-hole excitations $\ket{\Phi^{p\ldots}_{h\ldots}}$.
  }
\end{figure}

Instead of this naive approach, we can evolve the Hamiltonian operator, 
and achieve the desired decoupling without ever constructing a large matrix.
To this end, let us consider a representation of the Hamiltonian in a
basis built from a reference Slater determinant $\ket{\Phi}$ and its particle-hole
excitations, as shown in the left panel of Fig.~\ref{fig:sr_decoupling}. 
For simplicity, we show a two-body Hamiltonian, which can at most couple the 
reference state $\ket{\Phi}$ to 2p2h excitations, and in general $n$p$n$h to 
$(n\pm2)$p$(n\pm2)$h states, yielding the schematic band structure on display. 
We now want to continuously evolve this Hamiltonian to the shape shown 
in the right panel of Fig.~\ref{fig:sr_decoupling}. Inspecting the evolved
Hamiltonian more closely, we see that we aim to decouple the matrix element
$\matrixe{\Phi}{\HO(s)}{\Phi}$ from all excitations in the limit $s\to\infty$ --- if we succeed,
we will have extracted an eigenvalue of $\HO$, and our unitary transformation 
will provide us with a mapping of
the Slater determinant $\ket{\Phi}$ onto an exact eigenstate $\ket{\Psi}$
of $\HO(0)$, typically the ground state.

Since we are working with a reference state and its excitations, it is
convenient to represent the Hamiltonian in terms of normal-ordered operators.
We introduce the compact notation
\begin{equation}
  \AO^{i_1\ldots i_k}_{j_1\ldots j_k}\equiv
  \aaO_{i_1}\ldots\aaO_{i_k}\aO_{j_k}\ldots\aO_{j_1}\,
\end{equation}
for strings of creation and annihilation operators, and the normal ordered 
operators 
\begin{align}
  \nord{\AO^i_k} &= \AO^i_k - \rho^i_k\\
  \nord{\AO^{ij}_{kl}} &= \AO^{ij}_{kl} - \nord{\AO^{i}_{k}}\rho^{j}_{l} + \text{perm.} 
  - \rho^{i}_{k}\rho^{j}_{l}+\text{perm.}\\
  \ldots \notag\,
\end{align}
where $\rho^{i}_{k}=\matrixe{\Phi}{\AO^{i}_{k}}{\Phi}$ is the usual one-body
density matrix of the reference state (see \cite{Hergert:2016jk,Hergert:2017kx} 
for details). An intrinsic nuclear $A$-body Hamiltonian with NN and 3N interactions,
\begin{equation}
  \HO = \TO-\Tcm + \VO_2 + \VO_3
\end{equation}
can then be exactly rewritten as
\begin{align}
  \HO &= E 
        + \sum_{ij}f^{i}_{j}\nord{A^{i}_{j}} 
        + \frac{1}{4}\sum_{ijkl}\Gamma^{ij}_{kl}\nord{A^{ij}_{kl}}
        + \frac{1}{36}\sum_{ijklmn}W^{ijk}_{lmn}\nord{A^{ijk}_{lmn}}\,,
\end{align}
with the individual zero- through three-body contributions 
\begin{align}
  E &\equiv \left(1-\frac{1}{A}\right)\sum_{ab}t^{a}_{b}\rho^{a}_{b}
        + \frac{1}{4}\sum_{abcd}\left(\frac{1}{A}t^{ab}_{cd} + v^{ab}_{cd}\right)\rho^{ab}_{cd}
      + \frac{1}{36}\sum_{abcdef}v^{abc}_{def} \rho^{abc}_{def}\,,
      \label{eq:def_mr_E0}\\
  f^{i}_{j} &\equiv \left(1-\frac{1}{A}\right)t^{i}_{j} + \sum_{ab}\left(\frac{1}{A}t^{ia}_{jb} + v^{ia}_{jb}\right)\rho^{a}_{b}
  + \frac{1}{4}\sum_{abcd}v^{iab}_{jcd}\rho^{ab}_{cd}\,,\label{eq:def_mr_f}   \\
  \Gamma^{ij}_{kl} &\equiv \frac{1}{A}t^{ij}_{kl} + v^{ij}_{kl} + \sum_{ab}v^{ija}_{klb}\rho^{a}_{b}\,,\label{eq:def_mr_Gamma}\\
  W^{ijk}_{lmn}&\equiv v^{ijk}_{lmn}\,.\label{eq:def_mr_W}
\end{align}
Here, $\rho^{a\ldots}_{i\ldots}=\matrixe{\Phi}{\AO^{i_1\ldots i_k}_{j_1\ldots j_k}}{\Phi}$ 
denote the $k$-body density matrices of the reference state. For reasons that will
become clear below, we refrain from simplifying these expressions by using the factorization
of Slater determinant's density matrices. We note that due to
the normal ordering, the zero-, one-, and two-body coefficients all contain \emph{in-medium}
contributions from the three-body interaction, i.e., contractions of three-body 
matrix elements with the density matrices of the reference state. This explains
the name In-Medium SRG, and will play an important role when we introduce
truncations of the IMSRG flow equations below.

Using this definition of the Hamiltonian, we can use Wick's theorem to evaluate
the many-body matrix elements that couple $\ket{\Phi}$ to excitations, i.e., 
\begin{align}
  \matrixe{\Phi^p_h}{\nord{\HO}}{\Phi}&=\matrixe{\Phi}{\nord{\AO^h_p}\nord{\HO}}{\Phi}=f^p_h\\
  \matrixe{\Phi^{pp'}_{hh'}}{\nord{\HO}}{\Phi}&=\matrixe{\Phi}{\nord{\AO^{hh'}_{pp'}}\nord{\HO}}{\Phi}=\Gamma^{pp'}_{hh'}\,
\end{align}
(see \cite{Hergert:2016jk,Hergert:2017kx}).
These matrix elements define our off-diagonal Hamiltonian
\begin{equation}\label{eq:def_Hod}
  \Hod\equiv\sum_{ph}f^p_h\nord{\AO^p_h} + \frac{1}{4}\sum_{pp'hh'}\Gamma^{pp'}_{h'h}\nord{\AO^{pp'}_{hh'}} + \text{H.c.}\,,
\end{equation}
which can then be used in Eq.~\eqref{eq:def_Wegner_general} or another suitable
ansatz for the generator (see \cite{Hergert:2016jk,Hergert:2017kx} for details) 
to evolve the Hamiltonian to the desired form shown in the right panel of 
Fig.~\ref{fig:sr_decoupling}. Note that the elimination of $\Hod$ also removes
the couplings between general $n$p$n$h and $(n\pm2)$p$(n\pm2)$h states, i.e., the
outermost side-diagonal blocks.

\subsection{Correlated Reference States}
The discussion so far was based on the use of a Slater determinant 
as the reference state. On the one hand, such an independent-particle reference is convenient 
because it allows us to
distinguish particle and hole states in our single-particle basis, but on the other
hand, it means that the unitary transformation must encode all many-body correlations. The 
unitarity of the IMSRG transformation offers us more flexibility, because we can control 
to what extent correlations are described by either the Hamiltonian or the reference 
state. To see this, we consider the stationary Schr\"odinger equation and apply $\UO(s)$:
\begin{equation}
  \left[\UO(s)\HO\UUO(s)\right] \UO(s)\ket{\Psi_k} = E_k \UO(s) \ket{\Psi_k} \,.
\end{equation}
$\UO(s)$ shifts correlations from the wave function into the evolved, RG-improved
Hamiltonian $\HO(s)=\UO(s)\HO\UUO(s)$. In many-body calculations with this Hamiltonian, 
the many-body method now ``only'' needs to approximate $\UO(s)\ket{\Psi_k}$ rather than 
$\ket{\Psi_k}$ --- in the case discussed so far, even a Slater determinant would be 
sufficient because $\UO(s)$ or $\HO(s)$ describe the correlations\footnote{We also recall the 
evolution of the NN interaction in Sec.~\ref{sec:srg}, where the solution of the 
Schr\"odinger equation for the deuteron with an SRG-evolved Hamiltonian was a very 
simple two-nucleon wave function which nevertheless provides the correct deuteron 
ground-state energy.}. There can be situations in which it is worthwhile to use the
reference state to describe certain types of correlations, especially if it can 
capture them more efficiently than through the particle-hole type expansion
in which the IMSRG-evolved Hamiltonian is set up (see Sec.~\ref{sec:static}).

\begin{figure}[t]
  \begin{center}
    \includegraphics[width=0.9\textwidth]{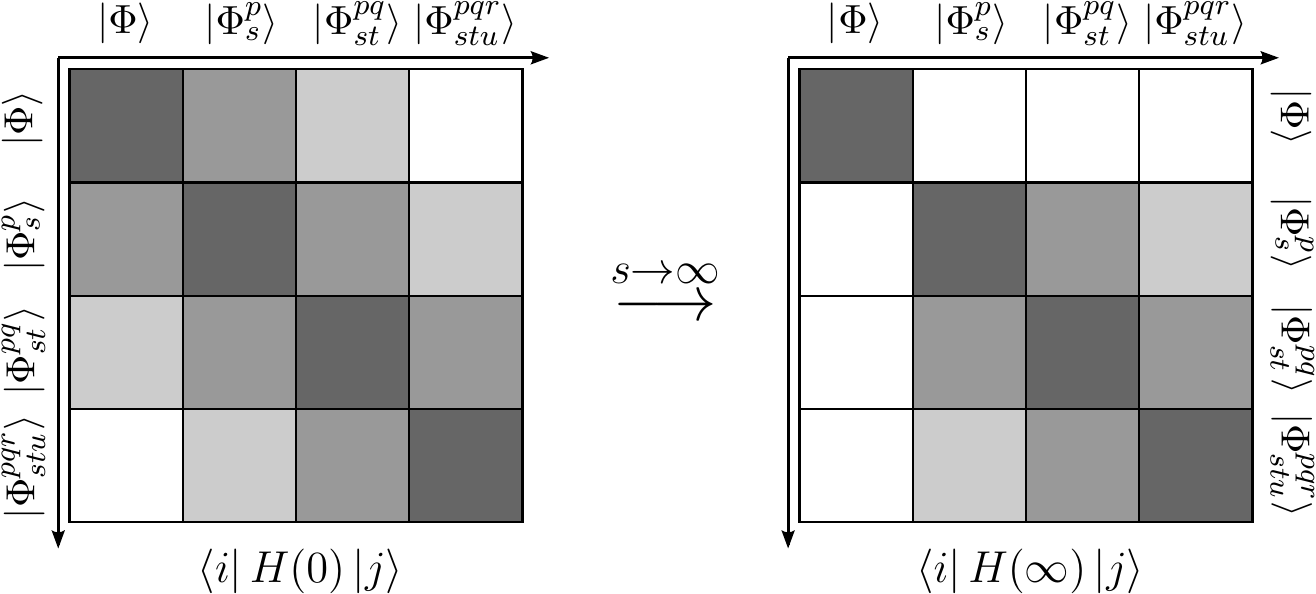}
  \end{center}
  \caption{\label{fig:mr_decoupling}
  Schematic view of MR-IMSRG decoupling in the many-body Hilbert space. $\ket{\Phi}$ denotes an arbitrary 
  reference state, and $\ket{\Phi^{p\ldots}_{s\ldots}}$ are suitably defined excitations. }
\end{figure}

The IMSRG framework can be extended to correlated reference states using
a generalized normal ordering developed by Kutzelnigg and Mukherjee 
\cite{Kutzelnigg:1997fk,Kong:2010kx}. It requires a new ingredient, the
\emph{irreducible $k$-body density matrices $\lambda^{(k)}$}. In the one-body 
case, this is just the usual density matrix
\begin{equation}
  \lambda^{i}_{k} \equiv \rho^i_k = \matrixe{\Phi}{\AO^i_k}{\Phi}\,,
\end{equation}
while for $k\geq 2$, 
\begin{align}
  \lambda^{ij}_{kl} &\equiv \rho^{ij}_{kl} - \AC\{\lambda^{i}_{k}\lambda^{j}_{l}\}\,, \label{eq:def_Lambda2}\\
  \lambda^{ijk}_{lmn}&\equiv \rho^{ijk}_{lmn} - \AC\{\lambda^{i}_{l}\lambda^{jk}_{mn}\} -
                      \AC\{\lambda^{i}_{l}\lambda^{j}_{m}\lambda^{k}_{n}\}\,,\label{eq:def_Lambda3}
\end{align}
etc., where $\AC\{\ldots\}$ fully antisymmetrizes the indices of the expression
within the brackets. The irreducible densities encode the correlation content of 
an arbitrary reference state $\ket{\Phi}$. If the reference state is a Slater 
determinant, i.e., an independent-particle state, the full two-body density matrix 
factorizes, and $\lambda^{(2)}$ vanishes (see, e.g., \cite{Hergert:2017kx}).

Normal-ordered one-body operators are constructed in the same manner as in
the standard normal ordering, e.g.,  
\begin{align}
  \nord{A^{a}_{b}}&= A^{a}_{b} - \lambda^{a}_{b}\,,\\
  \nord{A^{ab}_{cd}} &= A^{ab}_{cd}
                  - \lambda^{a}_{c}\nord{A^{b}_{d}} 
                  + \lambda^{a}_{d}\nord{A^{b}_{c}} 
                  - \lambda^{b}_{d}\nord{A^{a}_{c}}
                  + \lambda^{b}_{c}\nord{A^{a}_{d}}
                  - \lambda^{a}_{c}\lambda^{b}_{d}
                  + \lambda^{a}_{d}\lambda^{b}_{c}
                  - \lambda^{ab}_{cd}\,,\label{eq:nord_2B}\\
      \ldots &\notag
\end{align}
The expressions for the coefficients of the normal-ordered nuclear Hamiltonian,
Eqs.~\eqref{eq:def_mr_E0}--\eqref{eq:def_mr_W} remain unchanged, because we
wrote them in terms of the full density matrices. When we work with arbitrary reference 
states, the regular Wick's theorem extended with additional contractions like
{\setlength{\fboxsep}{1pt}
\begin{align}
  \nord{A^{\fbox{\scriptsize $ab$}}_{cd}}\nord{A^{ij}_{\fbox{\scriptsize $kl$}}} &= + \lambda^{\fbox{\scriptsize $ab$}}_{\fbox{\scriptsize $kl$}}\nord{A^{ij}_{cd}}\,,\label{eq:contract_lambda2Ba}\\ 
  \nord{A^{\fbox{\scriptsize $ab$}}_{c\fbox{\scriptsize$d$}}}\nord{A^{\fbox{\scriptsize $i$}j}_{\fbox{\scriptsize $kl$}}} &= - \lambda^{\fbox{\scriptsize $abi$}}_{\fbox{\scriptsize $dkl$}}\nord{A^{j}_{c}}\,,\label{eq:contract_lambda3B}\\ 
  \nord{A^{\fbox{\scriptsize $ab$}}_{\fbox{\scriptsize$cd$}}}\nord{A^{\fbox{\scriptsize $ij$}}_{\fbox{\scriptsize $kl$}}} &= +\lambda^{\fbox{\scriptsize $abij$}}_{\fbox{\scriptsize $cdkl$}}\,.\label{eq:contract_lambda4B}
\end{align}} 
(see \cite{Hergert:2017kx} for a detailed discussion). These contractions 
increase the number 
of terms when we expand operator products, e.g., in the analysis of many-body
matrix elements or the flow equations (see below), but fortunately, the overall 
increase in complexity is manageable.

In Fig.~\ref{fig:mr_decoupling}, we illustrate the schematic decoupling for the
Hamiltonian in a basis spanned by a correlated reference state and its excitations.
Since our reference state is no longer a Slater determinant, there are no well-defined
notions of a Fermi surface or particle and hole states any longer, and the indices
of the excitation operators run over all single-particle states
instead. Per construction, the reference state $\ket{\Phi}$ is still orthogonal to 
general excitations $\nord{\AO^{i_1\ldots i_k}_{j_1\ldots j_k}}\ket{\Phi}$ because
\begin{equation}
  \matrixe{\Phi}{\nord{\AO^{i_1\ldots i_k}_{j_1\ldots j_k}}}{\Phi}\,.
\end{equation}
The excitations themselves, however, are in general not orthogonal to each other,
and may even be linearly dependent. As a result, the structure of the evolved MR-IMSRG
Hamiltonian, shown in the right panel of Fig.~\ref{fig:mr_decoupling}, is not simplified 
as much as in the Slater determinant case, since matrix elements in the outer diagonals 
are not necessarily removed when novel contractions like \eqref{eq:contract_lambda2Ba}--\eqref{eq:contract_lambda4B} 
are in play. 

In the MR-IMSRG, the linear dependence implies that we are implicitly transforming a 
rank-deficient many-body Hamiltonian matrix and are possibly working harder than we 
need to by dragging along a space of spurious eigenstates with zero eigenvalues. Since 
the nuclear states we are interested in have absolute binding energies of several 
hundred MeVs, there has been no evidence that the presence of these spurious eigenstates
adversely affects the physical states in practical calculations.

\subsection{MR-IMSRG Flow Equations}
We are now ready to formulate the MR-IMSRG flow equations by applying the
generalized normal ordering and Wick's theorem to Eq.~\eqref{eq:opflow}. We express 
all operators in terms of normal-ordered strings of creation and annihilation operators. 
As discussed in Sec.~\ref{sec:srg}, each evaluation of the commutator in Eq.~\eqref{eq:opflow} 
induces operators of higher rank,
\begin{equation}\label{eq:induced}
  \comm{\nord{A^{ab}_{cd}}}{\nord{A^{ij}_{kl}}}= \delta_{ci}\nord{A^{abj}_{dkl}} +\ldots,
\end{equation}
hence we need to introduce truncations to obtain a computationally feasible 
many-body method. In contrast to Eq.~\eqref{eq:comm2B_vac}, we are now 
working with  \emph{normal-ordered} operators whose in-medium contributions 
have been absorbed into terms of lower rank and we expect the induced operators 
to be much weaker than in the free-space SRG case. There is ample empirical
evidence that the residual normal-ordered three-body piece can be neglected
to a very good approximation in nuclear physics \cite{Hagen:2007zc,Roth:2012qf,Hergert:2017kx}.
In our framework, we will eventually have to consider how the truncated
terms may feed back into operators of lower rank as we evolve the flow
equations (see below). In practice, we presently truncate all flowing operators 
at the two-body level to close the system of flow equations:
\begin{align} 
  \etaO(s) &\approx\etaO^{(1)}(s)+\etaO^{(2)}(s)\,,\label{eq:imsrg2_eta}\\
  \HO(s) &\approx E(s) + f(s) + \Gamma(s)\,,\label{eq:imsrg2_H}\\
  \totd{}{s}\HO(s) &\approx \totd{}{s}E(s) + \totd{}{s}f(s) + \totd{}{s}\Gamma(s)\label{eq:imsrg2_dH}\,.
\end{align}
This is the so-called MR-IMSRG(2) truncation \cite{Tsukiyama:2011uq,Tsukiyama:2012fk,Hergert:2013mi,Hergert:2013ij,Hergert:2014vn,Morris:2015ve,Hergert:2016jk}, which is a cousin to a variety of other truncated 
many-body schemes, in particular Coupled Cluster with Singles and Doubles excitations
(CCSD) \cite{Shavitt:2009,Hagen:2014ve,White:2002fk,Yanai:2006uq,Yanai:2007kx,
Nakatsuji:1976yq,Valdemoro:1987zl,Mukherjee:2001uq,Kutzelnigg:2002kx,Kutzelnigg:2004vn,Kutzelnigg:2004ys,Mazziotti:2006fk}, although the latter is based on non-unitary similarity transformations.

The MR-IMSRG(2) flow equations are given by \cite{Hergert:2013ij,Hergert:2014vn,Hergert:2017kx}
\begin{align}
  \totd{E}{s} &=     
    \sum_{ab}(n_{a}-n_{b})\eta^{a}_{b}f^{b}_{a}
    +\frac{1}{4}\sum_{abcd}
        \left(\eta^{ab}_{cd}\Gamma^{cd}_{ab}-\Gamma^{ab}_{cd}\eta^{cd}_{ab}\right)
        n_{a}n_{b}\bar{n}_{c}\bar{n}_{d}
    \notag\\
  &\hphantom{=}
    +\frac{1}{4}\sum_{abcd}\left(\totd{}{s}\Gamma^{ab}_{cd}\right)\lambda^{ab}_{cd}
    +\frac{1}{4}\sum_{abcdklm}\left(\eta^{ab}_{cd}\Gamma^{kl}_{am}-\Gamma^{ab}_{cd}\eta^{kl}_{am}\right)\lambda^{bkl}_{cdm}\,,
    \label{eq:mr_flow_0b_tens}
  \\[10pt]
  \totd{}{s}f^{i}_{j} &=
    \sum_{a}\left(\eta^{i}_{a}f^{a}_{j}-f^{i}_{a}\eta^{a}_{j}\right)
    +\sum_{ab}\left(\eta^{a}_{b}\Gamma^{bi}_{aj}-f^{a}_{b}\eta^{bi}_{aj}\right)(n_{a}-n_{b})
  \notag\\
  &\hphantom{=}
  +\frac{1}{2}\sum_{abc}
    \left(\eta^{ia}_{bc}\Gamma^{bc}_{ja}-\Gamma^{ia}_{bc}\eta^{bc}_{ja}\right)\left(n_{a}\bar{n}_{b}\bar{n}_{c}+\bar{n}_{a}n_{b}n_{c}\right)
  \notag\\
  &\hphantom{=}
      +\frac{1}{4}\sum_{abcde}\left(\eta^{ia}_{bc}\Gamma^{de}_{ja}-\Gamma^{ia}_{bc}\eta^{de}_{ja}\right)\lambda^{de}_{bc} 
    +\sum_{abcde}\left(\eta^{ia}_{bc}\Gamma^{be}_{jd}-\Gamma^{ia}_{bc}\eta^{be}_{jd}\right)\lambda^{ae}_{cd}
  \notag\\
  &\hphantom{=}
      -\frac{1}{2}\sum_{abcde}\left(\eta^{ia}_{jb}\Gamma^{cd}_{ae}-\Gamma^{ia}_{jb}\eta^{cd}_{ae}\right)\lambda^{cd}_{be}
    +\frac{1}{2}\sum_{abcde}\left(\eta^{ia}_{jb}\Gamma^{bc}_{de}-\Gamma^{ia}_{jb}\eta^{bc}_{de}\right)\lambda^{ac}_{de}\,,
  \label{eq:mr_flow_1b_tens}\\[10pt]
  \totd{}{s}\Gamma^{ij}_{kl}&=  
  \sum_{a}\left(\eta^{i}_{a}\Gamma^{aj}_{kl}+\eta^{j}_{a}\Gamma^{ia}_{kl}-\eta^{a}_{k}\Gamma^{ij}_{al}-\eta^{a}_{l}\Gamma^{ij}_{ka}
  -f^{i}_{a}\eta^{aj}_{kl}-f^{j}_{a}\eta^{ia}_{kl}+f^{a}_{k}\eta^{ij}_{al}+f^{a}_{l}\eta^{ij}_{ka}\right)
  \notag\\
  &\hphantom{=}
    +\frac{1}{2}\sum_{ab}\left(\eta^{ij}_{ab}\Gamma^{ab}_{kl}-\Gamma^{ij}_{ab}\eta^{ab}_{kl}\right)
     \left(1-n_{a}-n_{b}\right)
  \notag\\
  &\hphantom{=}
    +\sum_{ab}(n_{a}-n_{b})\left(\left(\eta^{ia}_{kb}\Gamma^{jb}_{la}-\Gamma^{ia}_{kb}\eta^{jb}_{la}\right)-\left(\eta^{ja}_{kb}\Gamma^{ib}_{la}-\Gamma^{ja}_{kb}\eta^{ib}_{la}\right)\right)\,,
  \label{eq:mr_flow_2b_tens}
\end{align}
where the $s$-dependence has been suppressed for brevity. We work in the natural 
orbital basis obtained by diagonalizing the one-body density matrix,
\begin{equation}
  \lambda^i_k=n_i\delta^i_k\,,
\end{equation}
where $0\leq n_i\leq 1$ indicates the occupation of the i-th orbital
in the correlated reference state. Analogously, $0\leq \nn_i=1-n_i\leq 1$
are the eigenvalues of the hole density matrix
\begin{equation}
  \xi^{i}_{j} \equiv \lambda^{i}_{j} - \delta^{i}_{j}\,.
\end{equation}

Note that due to the use of general reference states, the MR-IMSRG flow equations 
also include couplings to correlated pairs and triples of nucleons in the
reference state through the irreducible density matrices $\lambda^{(2)}$ and 
$\lambda^{(3)}$. The single-reference limit for a Slater determinant
reference state can be obtained by setting the irreducible density matrices 
$\lambda^{(2)}$ and $\lambda^{(3)}$ to zero, and noting that the occupation
numbers now become either 0 or 1.

Superficially, the computational cost for the evaluation of the MR-IMSRG(2) 
flow equations is dominated by the final term of Eq.~\eqref{eq:mr_flow_0b_tens},
which is of $\OC(N^7)$, where $N$ is the size of the single-particle basis.
However, most practical ans\"atze for correlated reference states impose
conditions on the density matrix that strongly limit the number of non-zero
matrix elements. For instance, the particle-number projected Hartree-Fock-Bogoliubov (HFB) 
\cite{Ring:1980bb} reference states used in our initial applications (see Sec.~\ref{sec:app}) have almost
diagonal $\lambda^{(k)}$ matrices. Moreover, relevant correlations are 
often restricted to a so-called valence or active space consisting of
a small number of single-particle states, so that only density matrix 
elements with \emph{all} indices belonging to this space are non-zero.
Thus, the main driver of the computational effort is the integration of the
two-body flow equation, at $\OC(N^6)$, which is the same naive scaling as
for CCSD and similar methods. For closed-shell references, we can exploit
the distinction of particle and hole indices to achieve exactly the same
scaling as for single-reference CCSD theory, namely $\OC(N_h^2N_p^4)$.

\section{Selected Applications\label{sec:app}}

\subsection{Reshuffling of Ground-State Correlations: $\nuc{Ca}{40}$}
\begin{figure}[t]
  \begin{center}
    \includegraphics[width=0.9\textwidth]{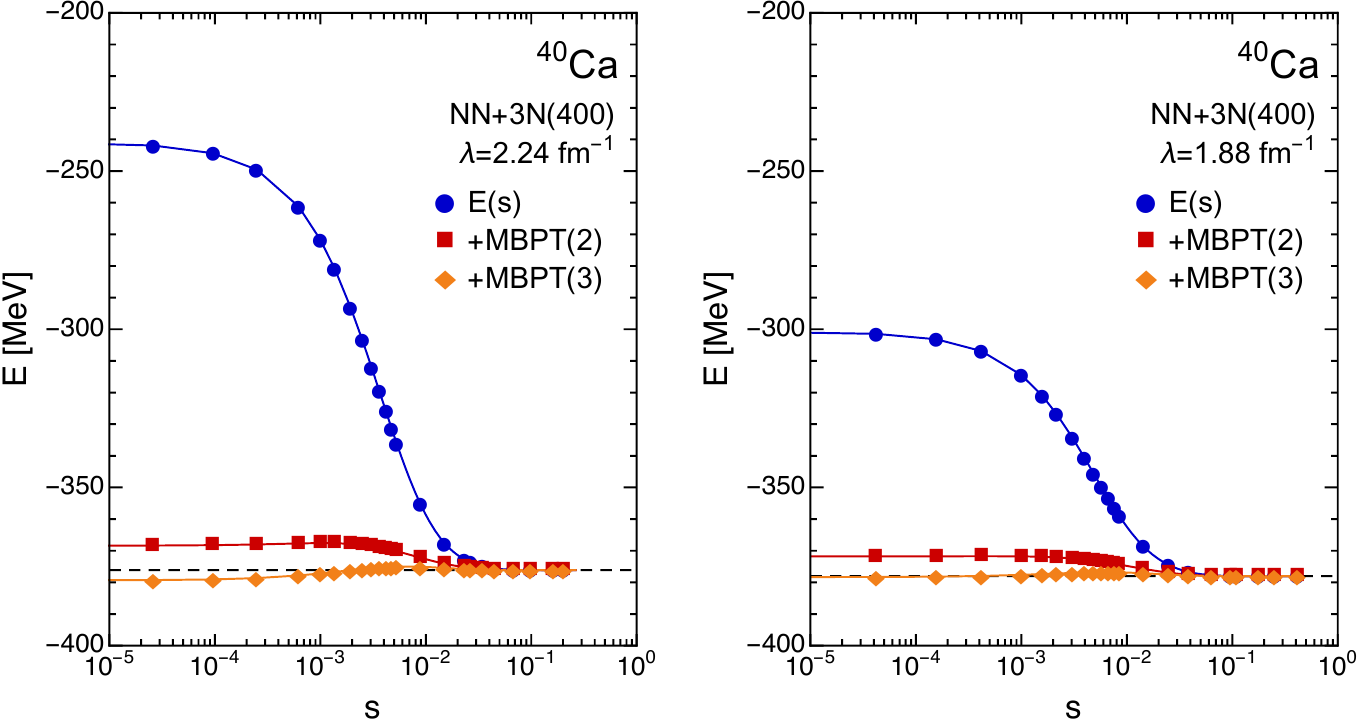}
  \end{center}
  \vspace{-10pt}
  \caption{\label{fig:mbpt}IMSRG(2) flow for $\nuc{Ca}{40}$ for a chiral NN+3N interaction,
    denoted NN+3N(400) in the following, at
    two different resolution scales $\lambdaSRG=2.24\fmi$ and $1.88\fmi$ (see Refs.~\cite{Hergert:2017kx,Hergert:2016jk}
    for details).
    We show the flowing ground-state energy $E(s)$, and the sum of $E(s)$ and perturbative energy
    corrections evaluated with the evolving Hamiltonian $\HO(s)$ to illustrate the re-shuffling
    of correlations into the Hamiltonian (see text). The dashed lines indicate the final IMSRG(2)
    energies.
  }
\end{figure}

To illustrate the effect of the (MR-)IMSRG flow in practice, we first consider
a ground-state calculation for the closed-shell nucleus $\nuc{Ca}{40}$. As our
reference, we use a Slater determinant that is optimized by performing a Hartree-Fock 
calculation with a chiral NN+3N interaction. Figure \ref{fig:mbpt} shows the IMSRG(2) 
ground-state energy $E(s)$ of $\nuc{Ca}{40}$ as a function of the flow parameter, which 
corresponds to the zero-body piece of the evolved normal-ordered Hamiltonian, 
Eq.~\eqref{eq:def_mr_E0}. As we integrate the flow equations to increasing $s$, the 
Hamiltonian is RG-improved with correlations that
are otherwise only accessible with post-Hartree Fock methods (see Refs.~\cite{Hergert:2016jk,Hergert:2017kx} for
details). This is evident
if we also consider the size of the second and third-order MBPT energy corrections, 
evaluated with $\HO(s)$. After a few integration 
steps, the energy corrections are absorbed into the Hamiltonian, which implies that
we could perform a Hartree-Fock calculation with $\HO(s)$ and obtain the 
fully correlated ground-state energy, up to truncation errors!

Figure \ref{fig:mbpt} also illustrates that the size of the MBPT corrections
depends significantly on the resolution scale of the Hamiltonian (see Sec.~\ref{sec:srg}).
At the higher resolution scale $\lambdaSRG=2.24\,\fmi$, we gain about $130\,\MeV$ 
of binding energy from correlations. At the lower resolution $\lambdaSRG=1.88\,\fmi$, 
the HF reference state is already significantly lower in energy, so the energy
gains from many-body correlations are less pronounced. This behavior is
expected as interactions become increasingly soft, and thereby more 
perturbative (see, e.g., \cite{Bogner:2010pq,Tichai:2016vl}). Note that the final
ground-state energies for $\lambdaSRG=2.24\,\fmi$ and $1.88\,\fmi$ are 
almost identical. As discussed in section \ref{sec:srg}, all results
should ideally be invariant under arbitrary changes of $\lambdaSRG$, which 
seems to be the case for the specific NN+3N Hamiltonian we used here
\cite{Roth:2011kx,Roth:2014fk}. 

\subsection{Ground-State Energies Along Isotopic Chains: The Oxygen Isotopes}

\begin{figure}[t]
  \begin{center}
    \includegraphics[width=0.5\textwidth]{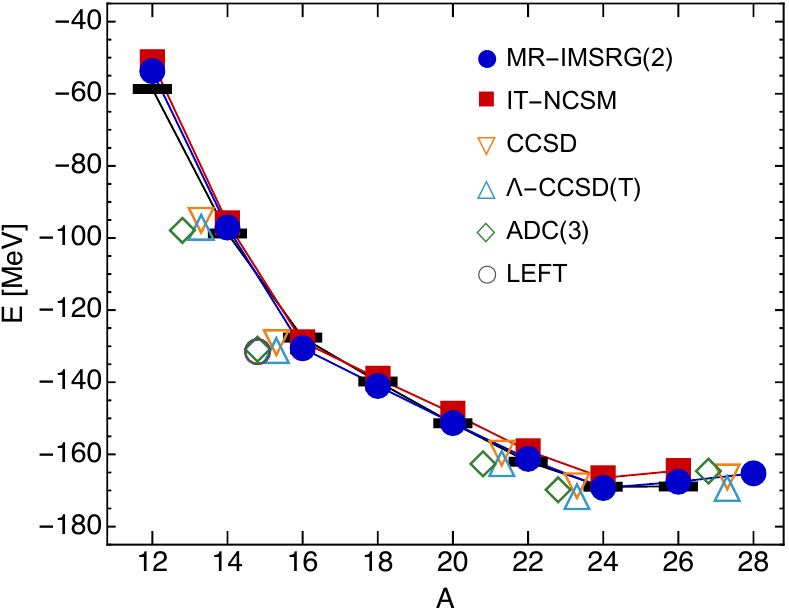}
  \end{center}
  \vspace{-10pt}
  \caption{\label{fig:OXX_methods} Ground-state energies of the oxygen isotopes from 
    MR-IMSRG(2) and other many-body approaches, using the NN+3N(400) interaction
    at $\lambdaSRG=1.88\fmi$. Some data points were offset horizontally
    to enhance the readability of the figure.  
    The ADC(3) Self-Consistent Green's Function results \cite{Cipollone:2013uq,Cipollone:2015fk}
    were obtained for $\lambdaSRG=2.0\,\fmi$, but the dependence of energies
    on $\lambdaSRG$ is very weak (see Fig.~\ref{fig:mbpt}). Black bars indicate experimental data \cite{Wang:2012uq}.
    }
\end{figure}

The oxygen isotopic chain is an important testing ground for \emph{ab initio}
nuclear structure theory because it is accessible to a wide array of exact
and approximate many-body methods
\cite{Otsuka:2010cr,Roth:2011kx,Hagen:2012oq,Hergert:2013ij,Cipollone:2013uq,Epelbaum:2014kx,Cipollone:2015fk,Holt:2013fk,Bogner:2014tg,Stroberg:2016fk}. As a bonus, the semi-magicity of the oxygen
isotopes --- the $Z=8$ protons are in a closed-shell configuration --- allows
us to exploit spherical symmetry in our calculations. In Fig.~\ref{fig:OXX_methods}, 
we compare MR-IMSRG(2) ground-state energies, obtained using particle-number projected
HFB vacua \cite{Hergert:2013ij,Hergert:2009zn,Ring:1980bb} as references for the 
open-shell nuclei, with a variety of other methods: 
\begin{itemize}
\item the Importance-Truncated No-Core Shell Model (IT-NCSM), which is an essentially 
exact diagonalization \cite{Roth:2009eu,Barrett:2013oq}, 
\item two types of Coupled Cluster, CCSD and $\Lambda$-CCSD(T), the latter of which 
includes approximate 3p3h (Triples) excitations \cite{Taube:2008kx,Taube:2008vn}, and
\item the ADC(3) Self-Consistent Green's Function (SCGF) approach \cite{Cipollone:2013uq,Cipollone:2015fk}
(obtained at a slightly different resolution, exploiting the approximate invariance
of the energies under variations of $\lambdaSRG$). 
\end{itemize}

Importantly, all used methods give consistent results when the same
input Hamiltonian is used, and the theoretical ground-state energies
agree within a few percent with experimental ground state energies. 
The systematically truncated methods, i.e., MR-IMSRG(2), CCSD, $\Lambda$-CCSD(T)
and ADC(3), agree very well with the exact IT-NCSM results, on the
level of 1\%--2\%, which provides us with an estimate of their truncation
errors. The MR-IMSRG(2) ground-state energy of $\nuc{O}{16}$ also agrees 
well with the result of a Nuclear Lattice EFT (NLEFT) calculation \cite{Epelbaum:2014kx}
directly using the chiral \NNLO{} Lagrangian.
Since the treatment of the nuclear many-body problem in NLEFT is completely 
different from all the other approaches compared here \cite{Lee:2009bh}, 
the good agreement is very encouraging.

The \emph{ab initio} calculations predict the neutron drip line at 
$\nuc{O}{24}$ in accordance with experimental findings \cite{Hoffman:2008ly,Caesar:2013qf}.
The addition of further neutrons no longer produces bound oxygen isotopes.
While absolute ground-state energies can change significantly under variations
of the resolution scale or other modifications of the initial Hamiltonian, the 
drip line signal turns out to be rather robust \cite{Hergert:2013ij}, if somewhat
exaggerated: In recent years, experimental studies have repeatedly revised the
energy of the $\nuc{O}{26}$ resonance downward \cite{Caesar:2013qf,Rogers:2015qf,Kohley:2015ay}.
Experimental data for $\nuc{O}{28}$ is forthcoming, which will most likely settle
the issue of the oxygen drip line. Still, the complex interplay of nuclear 
interactions, many-body and continuum effects causes the flat trend of experimental 
ground-state resonance energies beyond $\nuc{O}{24}$ that will preserve the
oxygen chain's status as an important testing ground for nuclear Hamiltonians 
and many-body methods for the foreseeable future.

\begin{figure}[t]
  \begin{center}
    \includegraphics[width=0.9\textwidth]{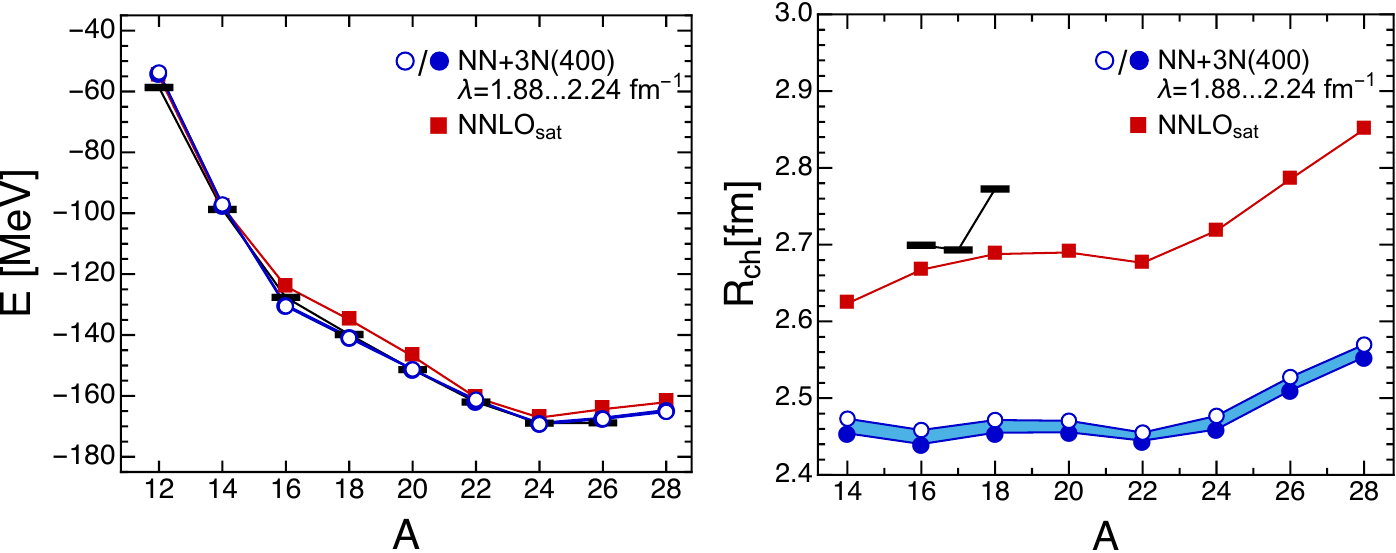}
  \end{center}
  \vspace{-10pt}
  \caption{\label{fig:OXX} MR-IMSRG(2) ground-state energies and charge radii of the oxygen 
    isotopes for \NNLOsat{} and NN+3N(400) at $\lambdaSRG=1.88,\ldots,2.24\fmi$ 
    (see text). Black bars indicate experimental 
    data \cite{Wang:2012uq,Angeli:2013rz}.
  }
\end{figure}

While chiral NN+3N interactions give a good reproduction of the oxygen
ground-state energies, their performance can differ significantly for other
observables, as illustrated in Fig.~\ref{fig:OXX}. The NN+3N(400) Hamiltonian \cite{Roth:2011kx,Gazit:2009qf}
that was also used in Figs.~\ref{fig:mbpt} and \ref{fig:OXX_methods} underestimates the 
experimentally known oxygen charge radii by about 10\%, and misses the sharp 
increase for $\nuc{O}{18}$. The other shown interaction, \NNLOsat{} 
\cite{Ekstrom:2015fk} improves the situation significantly, in part 
because its LECs (cf.~Sec.~\ref{sec:eft}) are optimized with a protocol
that includes selected many-body data, including the charge radius of
$\nuc{O}{16}$ --- however, it also falls short in $\nuc{O}{18}$ 
\cite{Lapoux:2016xu}. 

\begin{figure}[t]
  \begin{center}
    \includegraphics[width=\textwidth]{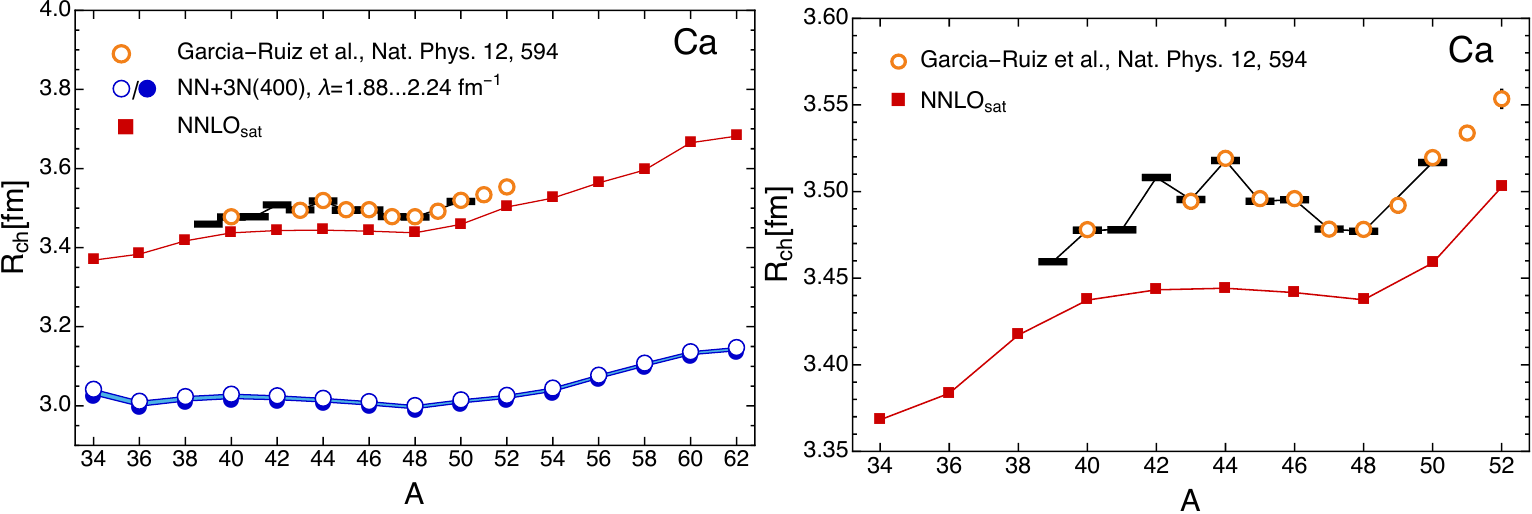}
  \end{center}
  \vspace{-10pt}
  \caption{\label{fig:CaXX} MR-IMSRG(2) charge radii of the calcium
    isotopes for \NNLOsat{} and NN+3N(400) with $\lambdaSRG=1.88,\ldots,2.24\fmi$. 
    Note the scale in the zoomed-in second panel. 
    Black bars indicate experimental data \cite{Wang:2012uq,Angeli:2013rz}.
  }
\end{figure}

As we go to heavier nuclei, the NN+3N(400) increasingly overestimates
the nuclear binding energy, up to about 1 MeV/nucleon in the light tin
isotopes, which are the heaviest nuclei whose ground states we can 
converge in the MR-IMSRG(2) scheme for this specific input. This overbinding then goes along
with an expected underestimation of nuclear charge radii, e.g., for the
calcium isotopes shown in Fig.~\ref{fig:CaXX}. \NNLOsat{}, on the other
hand, is the first chiral NN+3N interaction that gives realistic binding
energies and charge radii for $\nuc{Ca}{40}$ and $\nuc{Ca}{48}$ at the
same time \cite{Hagen:2015ve,Garcia-Ruiz:2016fk}. Unfortunately, it is
also a considerably harder interaction than NN+3N(400), and we already
encounter convergence problems if we move a few isotopic chains past
calcium.

Focusing on the calcium radii shown in the zoomed plot in the right panel
of Fig.~\ref{fig:CaXX}, we see that we do not reproduce the inverted 
parabolic trend between $\nuc{Ca}{40}$ and $\nuc{Ca}{48}$. In traditional
nuclear Shell Model/Configuration Interaction (CI) calculations with
empirical interactions, this is explained due to configuration mixing of
states containing 4p4h (Quadruples in chemistry parlance) and possibly 
higher excitations \cite{Caurier:2001oq} that are not included in the MR-IMSRG(2) 
due to the truncation of the operators at the two-body level,
and the use of spherical, particle-number projected HFB vacua as 
reference states in these calculations. We will discuss this issue in
more depth in the next section.

\section{\label{sec:static}Reference States with Static Correlation}

As explained in Sec.~\ref{sec:mrimsrg}, the IMSRG allows us to control
which correlations are built into the reference state and the RG-improved
Hamiltonian, respectively. This flexibility offers us an appealing option
for addressing the lack of specific features in nuclear charge radii that
we described at the end of Sec.~\ref{sec:app}.

Many-body bases built from a Slater determinant reference and its particle-hole
excitations work best for systems with large gaps in the single-particle 
spectrum, such as closed-shell nuclei. If the gap is small, particle-hole 
excited basis states can be near-degenerate with the reference determinant, 
which results in strong configuration mixing. When the mixing involves 
configurations in which many nucleons are excited simultaneously, we will have
\emph{static} or \emph{collective correlations} in the wave function, as opposed 
to \emph{dynamic correlations} that result from a small number of nucleons only. 

Important examples are the emergence of nuclear superfluidity \cite{Dean:2003ei}
or diverse rotational and vibrational bands in open-shell nuclei (see, e.g., \cite{Bohr:1999vn}).
These phenomena are conveniently described by using the concept of intrinsic
wave functions that explicitly break appropriate symmetries of the Hamiltonian.
For instance, nuclear superfluidity can be treated to leading-order by 
introducing quasiparticles that are defined as superpositions of particle
and holes \cite{Bardeen:1957tv,Bogoliubov:1958ci,Ring:1980bb}. The ground-state of an open-shell nucleus can be expressed as
an antisymmetrized product state of these quasiparticles, which can then
be optimized variationally --- this is precisely the HFB method that was the first step in the generation of references states 
for the studies of isotopic chains discussed in Sec.~\ref{sec:app}. The intrinsic 
HFB wave functions are superpositions of states with different particle numbers,
hence we eventually proceeded to restore the symmetry by particle number projection.
Such approaches that involve symmetry breaking and subsequent restoration via 
projection have a long history in nuclear many-body theory
\cite{Peierls:1973fk,Ring:1980bb,Egido:1982sd,Robledo:1994qf,Flocard:1997fx,Sheikh:2000xx,Dobaczewski:2007hh,Bender:2009rv,Duguet:2009ph,Lacroix:2009aj,Lacroix:2012vn,Duguet:2015ye}). 

\begin{figure}[t]
  \begin{center}
    \includegraphics[width=\textwidth]{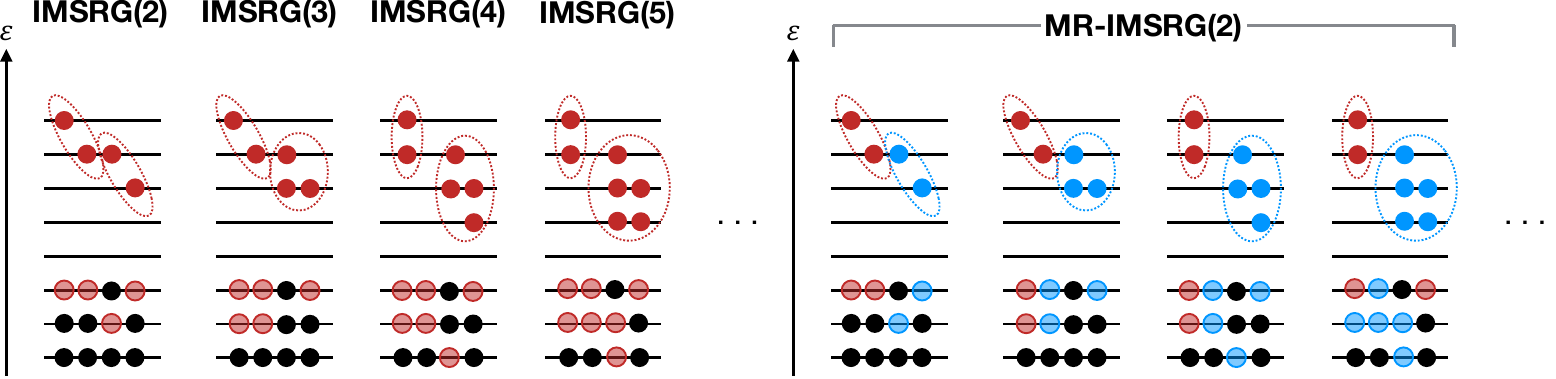}
  \end{center}
  \vspace{-10pt}
  \caption{\label{fig:static_corr}
    Schematic view of correlations in nuclei. Solid
    circles indicate nucleons, transparent circles hole states, and dashed
    ellipses indicate correlations between nucleons. The left panel
    indicates the IMSRG truncation that would be required to describe the 
    sketched 2p2h (Doubles), 3p3h (Triples), etc. correlations. In the right
    panel, certain 2p2h as well as all 3p3h and higher correlations (indicated 
    in blue) are built into a correlated
    wave function that then serves as the reference state for an MR-IMSRG(2)
    calculation.
  }
\end{figure}

The benefit of symmetry breaking and projection methods is that they 
capture static correlations in a set of simple one-body transition density 
matrices \cite{Ring:1980bb,Duguet:2015ye}. Instead of using a Slater determinant 
reference state and trying to describe static correlations via 3p3h (Triples),
4p4h (Quadruples) and higher excitations of this reference via a high-rank IMSRG 
evolution (left panel of Fig.~\ref{fig:static_corr}),
we can use a correlated reference state whose irreducible density matrices
$\lambda^{(k)}$ we can construct on the fly with little effort, and perform
an MR-IMSRG(2) evolution to treat the dynamical correlation on top of this
state (right panel of Fig.~\ref{fig:static_corr}).

\begin{figure}[t]
  \begin{center}
    \includegraphics[width=0.5\textwidth]{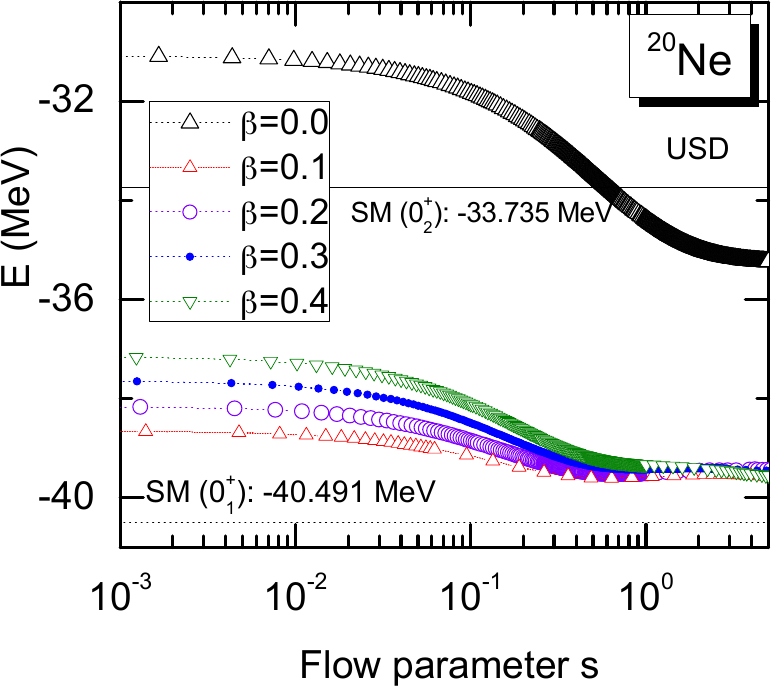}
  \end{center}
  \vspace{-10pt}
  \caption{\label{fig:Ne20}
    MR-IMSRG(2) energy of $\nuc{Ne}{20}$, starting from reference states with 
    prolate intrinsic deformations $\beta\geq0$ that are subsequently projected
    on angular momentum $J^\pi=0^+$. These proof-of-principle calculations were
    performed with the USD Shell Model interaction \cite{Wildenthal:1984qf,Brown:1988xy}.
  }
\end{figure}

As a proof of principle, we show MR-IMSRG(2) calculations for the nucleus
$\nuc{Ne}{20}$ in Fig.~\ref{fig:Ne20}. Both the proton and neutron shells
are open in $\nuc{Ne}{20}$, which causes the ground-state to develop prolate 
intrinsic deformation. We obtain reference states by performing HFB calculations 
with a constraint on the quadrupole moment of the many-body state, and subsequently
projecting them on good particle number and angular momentum $J^\pi=0^+$. 
The references with prolate deformation ($\beta>0.0$) flow towards 
the same $0^+$ ground state whose energy is close to the result of an exact 
diagonalization with the empirical USD Hamiltonian \cite{Wildenthal:1984qf,Brown:1988xy} we use in our calculation. A spherical reference state ($\beta=0.0$) flows to an excited
$0^+$ state in the spectrum of $\nuc{Ne}{20}$ instead.

\begin{figure}[t]
  \begin{center}
    \includegraphics[width=\textwidth]{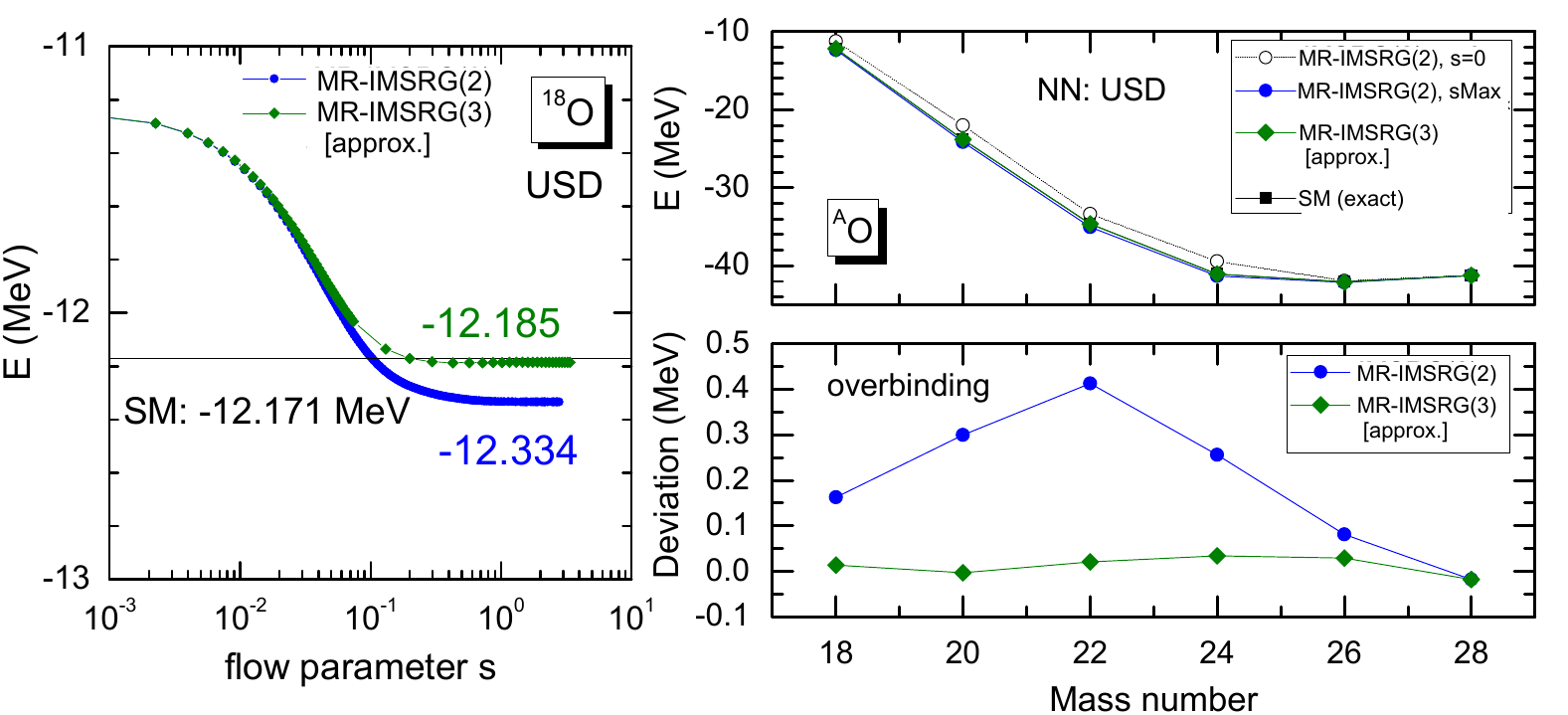}
  \end{center}
  \vspace{-10pt}
  \caption{\label{fig:OXX_imsrg3}
    MR-IMSRG(2) and approximate MR-IMSRG(3) ground-state energies of the oxygen
    isotopes, using the USD Shell Model interaction \cite{Wildenthal:1984qf,Brown:1988xy}. 
  }
\end{figure}

The discrepancies between the MR-IMSRG(2) energies for the intrinsically 
prolate and spherical states and the exact Shell Model results are caused by
the truncation of the flow equations at the two-body level. While a full
MR-IMSRG(3) calculation would have a computational cost of $\OC(N^9)$,
one-step approximations that leverage information from the MR-IMSRG(2)
can be implemented at a more manageable $\OC(N^7)$ effort \cite{Morris:2016xp}, 
analogous to completely renormalized Coupled Cluster schemes that account 
for leading Triples excitations \cite{Piecuch:2005dp,Binder:2013fk}. 
In the left panel of Fig.~\ref{fig:OXX_imsrg3}, we perform such an approximate
MR-IMSRG(3) calculation for $\nuc{O}{18}$, and find that we 
recover the exact diagonalization result\cite{Brown:2006fk}. In the right 
panel, we show that the same improvement is found for all even oxygen isotopes, which
is very promising for future applications of the approximate MR-IMSRG(3)
in general open-shell nuclei.

\section{\label{sec:heff}RG-Improved Hamiltonians: Merging (MR-)IMSRG with Other Many-Body Methods}
In the previous sections we have focused on using the (MR-)IMSRG to
directly calculate the ground-state energy of nuclei, expressed as the zero-body part of the
RG-improved Hamiltonian in the limit $s\to\infty$. We are only realizing
a fraction of the power and utility of the (MR-)IMSRG framework in this
way: In principle, we have the entire Hamiltonian at our disposal, and
not just at the end-point of the flow, but along an entire RG trajectory,
which equips us with additional diagnostic power --- we recall the variation 
of the triton ground-state energy with the resolution scale in a free-space 
SRG, and how we can draw conclusions about missing induced forces from it
(see Sec.~\ref{sec:srg}). 

In our discussion of Fig.~\ref{fig:sr_decoupling}, we pointed out that 
the suppression of the off-diagonal matrix elements that couple the ground
state to 2p2h excitations will also eliminate the outermost side diagonals
of the Hamiltonian matrix, which implies that it can be beneficial to use
the RG-improved Hamiltonian as an input for a secondary many-body calculation.
We have explored this idea in several ways: 

\begin{itemize}
\item With a simple modification of the definition of the off-diagonal Hamiltonian,
it is possible to systematically derive effective interactions and operators for 
valence-space CI calculations, i.e., the traditional interacting
Shell Model \cite{Tsukiyama:2012fk,Bogner:2014tg,Stroberg:2016fk,Stroberg:2017sf}.
This allows us to make use of existing technology to study any nucleus whose 
low-lying states can be converged in typical Shell Model calculations, explaining
the coverage of the nuclear chart we have shown in Fig.~\ref{fig:chart2017}. it
also directly connects Shell Model phenomenology to the underlying chiral interactions ---
the valence-space diagonalization is merely an approximation to a no-core,
large-scale diagonalization.
In applications for $sd$-shell nuclei ($8\leq Z\leq 20$), we were able to achieve
a surprisingly good agreement with calculations using the gold standard empirical
USD interaction \cite{Wildenthal:1984qf,Brown:1988xy}, which itself has an
rms deviation of a mere $\sim130\keV$ for more than 600 energy levels in this
region of the nuclear chart. 
\item The (MR-)IMSRG was recently merged with the No-Core Shell Model (NCSM)
into the so-called In-Medium NCSM, in an effort to combine the strengths of the
two approaches \cite{Gebrerufael:2017fk}. A NCSM calculation in a small space provides 
a correlated reference state for the normal ordering and MR-IMSRG evolution, 
which enhances the Hamiltonian with dynamical correlation corresponding whose 
description would otherwise require extremely large NCSM model spaces. For NCSM 
reference states, the decoupling of the ground-state from excitations is not 
perfect, hence a secondary small NCSM diagonalization is performed to extract 
the low-lying eigenvalues and the associated eigenstates. In the IM-NCSM, we
make explicit use of the Hamiltonian's flow-parameter dependence for diagnostic
purposes (see Ref.~\cite{Gebrerufael:2017fk} for more details).
\item Finally, the ground-state decoupled Hamiltonian can be used as input for
Equation-of-Motion (EOM) methods like the well-known Tamm-Dancoff (TDA) and Random 
Phase Approximations (RPA). TDA and RPA will actually become identical for an IMSRG
Hamiltonian, because the ground-state decoupling removes the feedback
of RPA correlations into the ground state \cite{Parzuchowski:2017yq}. EOM calculations
are complementary to the CI-based approaches discussed above because they allow us
to explore the effects of large single-particle bases, at the cost of limiting the
particle rank of the excitation operator for computational efficiency. Thus, they
are best suited for the description of dynamical correlations in excited states,
although the development of an MR-EOM approach based on correlated reference states
is in progress. 
\end{itemize}

The combination of the (MR-)IMSRG with CI/Shell Model techniques allows
us to treat both dynamical and static correlation (see Sec.~\ref{sec:static}),
but limits us through the factorial computational scaling of the
diagonalization step. As already pointed out by White in Ref.~\cite{White:2002fk},
the Density Matrix Renormalization Group (DMRG) can be an appealing 
and efficient alternative to large-scale CI methods, especially when
combined with (MR-)IMSRG decoupling (or Canonical Diagonalization in
his choice of words). This idea was pursed by Yanai and co-workers
in quantum chemistry \cite{Yanai:2006uq,Yanai:2007kx,Neuscamman:2009ve,Neuscamman:2010fk,Yanai:2010ys}),
and we are exploring its applications in nuclear physics. 

Finally, we are looking ahead at applications of the IMSRG framework
to the description of nuclei with cluster structures, and eventually
reaction processes. Such structures and processes are most conveniently modeled
on spatial lattices, which is why we are working on formulating the
IMSRG in that context. Tensor RG methods 
that generalize the ideas of DMRG to higher dimensions have been very 
successful in describing correlated systems on spatial lattices 
could be an ideal complement to the IMSRG (see, e.g., \cite{Vidal:2007if,Evenbly:2009cy,Evenbly:2015ht})
. Interestingly, these Tensor
RGs rely on unitary transformations to disentangle long- and short-range
physics that are at least superficially reminiscent of SRG and IMSRG
transformations --- this certainly merits further
investigation.

\section{Conclusions and Outlook}
In this contribution, we have attempted to illustrate the current reach and 
capabilities of modern \emph{ab initio} nuclear structure theory, using the
(MR-)IMSRG as a representative example. The discussion is by no means exhaustive, 
given the rapid rate at which new advances are being introduced. While our
focus was on the many-body method, the presented results show that the input
NN+3N interactions from chiral EFT are currently the dominant source of 
uncertainty in our theoretical results, but promising new chiral 
interactions \cite{Ekstrom:2015fk,Reinert:2017fv,Lynn:2016ec,Lynn:2017eu,Entem:2015qf,Entem:2015hl}
are now being explored by the nuclear many-body community. The MR-IMSRG will
be particularly useful in the community's efforts to confront these new 
Hamiltonians with current and forthcoming experimental data for medium-mass
and heavy open-shell nuclei.

Work is currently underway to extend the MR-IMSRG framework in several directions,
including the coupling to continuum degrees of freedom, and the treatment of 
intrinsic deformation. The present contribution showed first proof-of-principle
results for the latter, implementing the MR-IMSRG transformation with an 
intrinsically deformed, angular-momentum projected reference state. We
plan to extend this work further by using the Generator Coordinate Method
to mix multiple projected configurations (see, e.g., \cite{Ring:1980bb}).

Motivated by our successes in using the (MR-)IMSRG to construct RG-improved
Hamiltonians that can serve as input for the traditional nuclear Shell
Model/CI \cite{Bogner:2014tg,Stroberg:2016fk,Stroberg:2017sf}, 
the No-Core Shell Model \cite{Gebrerufael:2017fk}, and EOM methods
\cite{Parzuchowski:2017yq}, we seek to integrate the (MR-)IMSRG 
evolution with further approaches, in particular DMRG, which presents 
an appealing alternative to costly diagonalization 
methods for treating static correlations in configuration space.
Such a combination of the IMSRG for the Hamiltonian with an entanglement-based RG for 
wave functions could also offer interesting opportunities for describing 
strongly interacting systems on spatial or spacetime lattices. 

\section*{Acknowledgments}
We are indebted to our collaborators on IMSRG projects, namely E.~Gebrerufael, 
M.~Hjorth-Jensen, J.D.~Holt, R.~Roth, A.~Schwenk, S.R.~Stroberg and K.~Vobig,
as well as K.~Tsukiyama and F.~Yuan.
We also thank C.~Barbieri, S.~Binder, A.~Calci, T.~Duguet, F.~Evangelista, 
R.~J.~Furnstahl, G.~Hagen, K.~Hebeler, G.~R.~Jansen, J.~Simonis, V.~Som\`{a}, 
and K.~A.~Wendt for many useful discussions, as well as providing interaction
elements and/or theoretical results for comparison.

This publication is based on work supported in part by the National Science 
Foundation under Grants No.~PHY-1404159, PHY-1614130, and PHY-1713901, as well
as the U.S. Department of Energy, Office of Science, Office of Nuclear Physics 
under Grants No. DE-SC0008511 and DE-SC0008641 (NUCLEI SciDAC Collaboration), 
DE-FG02-97ER41019 and DE-SC0004142.

Computing resources were provided by the Michigan State University Institute 
for Cyber-Enabled Research (iCER), and the National Energy Research Scientific 
Computing Center (NERSC), a DOE Office of Science User Facility supported by 
the Office of Science of the U.S.~Department of Energy under Contract No. DE-AC02-05CH11231.


\bibliography{2017_mbt19_hergert.bib}


\end{document}